\documentclass[aps,pre,reprint, amsmath, amssymb,superscriptaddress]{revtex4-1}
\usepackage{graphicx} 
\usepackage{braket}
\usepackage[colorlinks,linkcolor=red,urlcolor=blue,citecolor=blue]{hyperref}
\usepackage{xcolor}
\usepackage[bb=boondox]{mathalfa}

\begin{document}
\title{Quantum stochastic thermodynamics in the mesoscopic-leads formulation}
\author{Laetitia P. Bettmann}
\email{bettmanl@tcd.ie}
\affiliation{School of Physics, Trinity College Dublin, College Green, Dublin 2, D02K8N4, Ireland}
\author{Michael J. Kewming}
\email{kewmingm@tcd.ie}
\affiliation{School of Physics, Trinity College Dublin, College Green, Dublin 2, D02K8N4, Ireland}
\author{Gabriel T. Landi}
\email{glandi@ur.rochester.edu}
\affiliation{Department of Physics and Astronomy, University of Rochester, Rochester, New York 14627, USA}
\author{John Goold}
\email{gooldj@tcd.ie}
\affiliation{School of Physics, Trinity College Dublin, College Green, Dublin 2, D02K8N4, Ireland}
\affiliation{Trinity Quantum Alliance, Unit 16, Trinity Technology and Enterprise Centre, Pearse Street, Dublin 2, D02YN67, Ireland}
\author{Mark T. Mitchison}
\email{mark.mitchison@tcd.ie}
\affiliation{School of Physics, Trinity College Dublin, College Green, Dublin 2, D02K8N4, Ireland}
\affiliation{Trinity Quantum Alliance, Unit 16, Trinity Technology and Enterprise Centre, Pearse Street, Dublin 2, D02YN67, Ireland}
\begin{abstract}
We introduce a numerical method to sample the distributions of charge, heat, and entropy production in open quantum systems coupled strongly to macroscopic reservoirs, with both temporal and energy resolution and beyond the linear-response regime.  
Our method exploits the mesoscopic-leads formulation, where macroscopic reservoirs are modeled by a finite collection of modes that are continuously damped toward thermal equilibrium by an appropriate Gorini-Kossakowski-Sudarshan-Lindblad master equation. Focussing on non-interacting fermionic systems, we access the time-resolved full counting statistics through a trajectory unraveling of the master equation. We show that the integral fluctuation theorems for the total entropy production, as well as the martingale and uncertainty entropy production, hold. Furthermore, we investigate the fluctuations of the dissipated heat in finite-time information erasure. Conceptually, our approach extends the continuous-time trajectory description of quantum stochastic thermodynamics beyond the regime of weak system-environment coupling.
\end{abstract}
\date{\today}

\maketitle
\section{Introduction}
Stochastic thermodynamics, initially formulated for classical systems \cite{seifert_stochastic_2012,ciliberto_experiments_2017, seifert_stochastic_2018,seifert_entropy_2005, lebowitz_gallavotticohen-type_1999, jarzynski_equalities_2011} and later extended to the quantum domain \cite{horowitz_quantum-trajectory_2012, hekking_quantum_2013, manzano_quantum_2022, manzano_quantum_2019, manzano_quantum_2022, goold_role_2016}, allows the description of energy transfer and entropy production along single trajectories of systems undergoing non-equilibrium processes.
Based on this framework, several significant insights into the second law of thermodynamics emerged. Notably, universal relations governing the statistics of fluctuating thermodynamic currents, referred to as fluctuation theorems, have been uncovered \cite{sekimoto_stochastic_nodate, seifert_stochastic_2012, jarzynski_equalities_2011, esposito_nonequilibrium_2009, campisi_colloquium_2011, deffner_nonequilibrium_2011, morikuni_quantum_2011, funo_quantum_2015, manzano_quantum_2018, bartolotta_jarzynski_2018, batalhao_experimental_2014, an_experimental_2015}. 

Our focus lies in exploring thermodynamic properties within scenarios where a central system is driven out of equilibrium due to its interaction with thermodynamic reservoirs or an external drive, leading to the generation of fluctuating currents carrying particles and heat. For open quantum systems, boundary effects are often comparable in magnitude to internal interactions and the coupling between the central system and reservoirs may be strong \cite{talkner_colloquium_2020}. In this context, the spectral properties of the reservoirs, which can be non-trivial, may play a crucial role for the dynamics of the central system.

Only a handful of methods exist that can address fluctuations of thermodynamic quantities in this far-from-equilibrium, strongly coupled setting. Non-equilibrium Green functions (NEGF) and scattering theory~\cite{blanter_shot_2000,esposito_nonequilibrium_2009,Agarwalla2012,Esposito2015} are generally limited to slowly driven or weakly interacting systems~\cite{Moskalets2004,Moskalets2014,Potatina2021}, while path integral methods~\cite{Aurell2018,Funo2018,Funo2018a} have been applied to numerically compute heat fluctuations in small, nonintegrable open quantum systems~\cite{Kilgour2019,Popovic2021} but scaling this to larger systems, e.g.~spin chains~\cite{Fux2023}, remains an open challenge. Another approach to strong-coupling thermodynamics exploits a Markovian embedding~\cite{woods_mappings_2014} such as the reaction-coordinate method~\cite{IlesSmith2014,Strasberg2016,Newman2017,Anto-Sztrikacs2021, Diba2023}, where a non-Markovian open quantum system is incorporated into a larger Markovian one, leading to equivalent dynamics for the original open system. A powerful many-body extension of this approach is the mesoscopic-leads formulation of quantum transport~\cite{subotnik_nonequilibrium_2009, lacerda_entropy_2023, lacerda_quantum_2022, gruss_landauers_2016, guimaraes_nonequilibrium_2016, elenewski_communication_2017, uzdin_markovian_2018, reichental_thermalization_2018, chen_markovian_2019, brenes_tensor-network_2020, dzhioev_super-fermion_2011, ajisaka_nonequlibrium_2012, ajisaka_nonequilibrium_2013, zelovich_state_2014, chen_simple_2014, oz_numerical_2020, schwarz_lindblad-driven_2016, schwarz_nonequilibrium_2018, lotem_renormalized_2020, elenewski_performance_2021, wojtowicz_dual_2021}, where macroscopic reservoirs are approximated by a finite number of damped modes in the same spirit as the pseudomode approach to open quantum systems~\cite{imamoglu_stochastic_1994,garraway_decay_1997, garraway_nonperturbative_1997, Tamascelli2018, Lambert2019, Mascherpa2020}. The mesoscopic-leads formalism has recently been adapted to compute fluctuating charge currents in driven systems~\cite{brenes_particle_2023}, but this method is not applicable to heat currents. Importantly, moreover, all the aforementioned approaches are tailored to compute the characteristic function, from which features of the probability distribution (e.g.~of charge, heat etc.) can only be extracted by numerical differentiation or Fourier transform, with their associated errors and limitations. 

In this work, we introduce a method to directly sample the probability distributions of charge, heat, and entropy production in driven, strongly coupled open quantum systems far from equilibrium. Our approach exploits a mesoscopic-leads setup where the dynamics of the extended system is described by a Gorini-Kossakowski-Sudarshan-Lindblad (GKSL) master equation (ME) \cite{lindblad_generators_1976,gorini_completely_1976}. Monitoring the individual exchanges of energy quanta with the reservoirs results in a stochastic unravelling of the dynamics into quantum-jump trajectories. Each trajectory represents the state of the extended system conditioned on the measurement record~\cite{wiseman_quantum_2009}, and sampling such trajectories recovers the full counting statistics of currents and other observables~\cite{Landi2024}. In particular, we show that trajectory sampling enables reconstruction of the full distributions of heat, work, and entropy production, with both temporal and energy resolution. We focus on non-interacting fermionic systems, although our methods could be combined with tensor-network methods~\cite{brenes_tensor-network_2020,lacerda_entropy_2023} in the future to address the stochastic thermodynamics of arbitrary interacting systems. In this context, we demonstrate thermodynamic consistency of this method at the stochastic level by verifying three fluctuation theorems for entropy production: specifically, for the total, uncertainty, and martingale entropy production~\cite{manzano_quantum_2022, manzano_quantum_2019}. As an application, we then study the full heat statistics of information erasure in a driven quantum dot coupled strongly to a fermionic reservoir. Our approach enables us to access the full heat distribution for arbitrary driving speeds and coupling strengths, going beyond previous work on finite-time quantum information erasure~\cite{miller_quantum_2020,Vu2022,rolandi_finite-time_2022}.

We emphasise that, to our knowledge, no other existing method allows for direct stochastic sampling of thermodynamic quantities in non-Markovian settings. Having access to individual trajectories not only gives additional insight beyond analysis of the cumulants, e.g.~when examining the physical processes underpinning rare fluctuations~\cite{miller_quantum_2020}, but also unlocks the toolkit of continuous measurement theory for investigating measurement and feedback in non-Markovian quantum many-body systems. 
The paper is structured as follows: in Sec.~\ref{sec: mesoleads_formulation} the mesoscopic-leads formulation for fermions is introduced. Then, in Sec.~\ref{sec: state_dynamics} the unconditional state dynamics is discussed, and further specified for non-interacting systems (Sec.~\ref{sec: state_dynamics_gauss}). In Sec.~\ref{sec: cond_evo} the conditional dynamics is presented. In Sec.~\ref{sec: cond_thermo}, we discuss the thermodynamically consistent inference of particle (Sec.~\ref{sec: cond_JP}) and energy current (Sec.~\ref{sec: cond_JE}) along single trajectories, and the energy's splitting into heat and work contributions (Sec.~\ref{sec: meas_work}). 
In Sec.~\ref{sec: EP_TPM}, we discuss the total entropy production along a trajectory, relying on a two-point measurement scheme (TPM) in the mesoscopic-leads formalism. 
We demonstrate the emergence of integral fluctuation theorems for both the total entropy production and the uncertainty and martingale entropy production separately, in Sec.~\ref{sec: val_FT}. Subsequently, we study the fluctuations of dissipated heat during finite-time information erasure in Sec.~\ref{sec: heat_ft_le}.

\section{Mesoscopic-leads formulation}
\label{sec: mesoleads_formulation}
\subsection{State dynamics}
\label{sec: state_dynamics}
\begin{figure}[t]
\begin{center}
\includegraphics[width=0.6\linewidth]{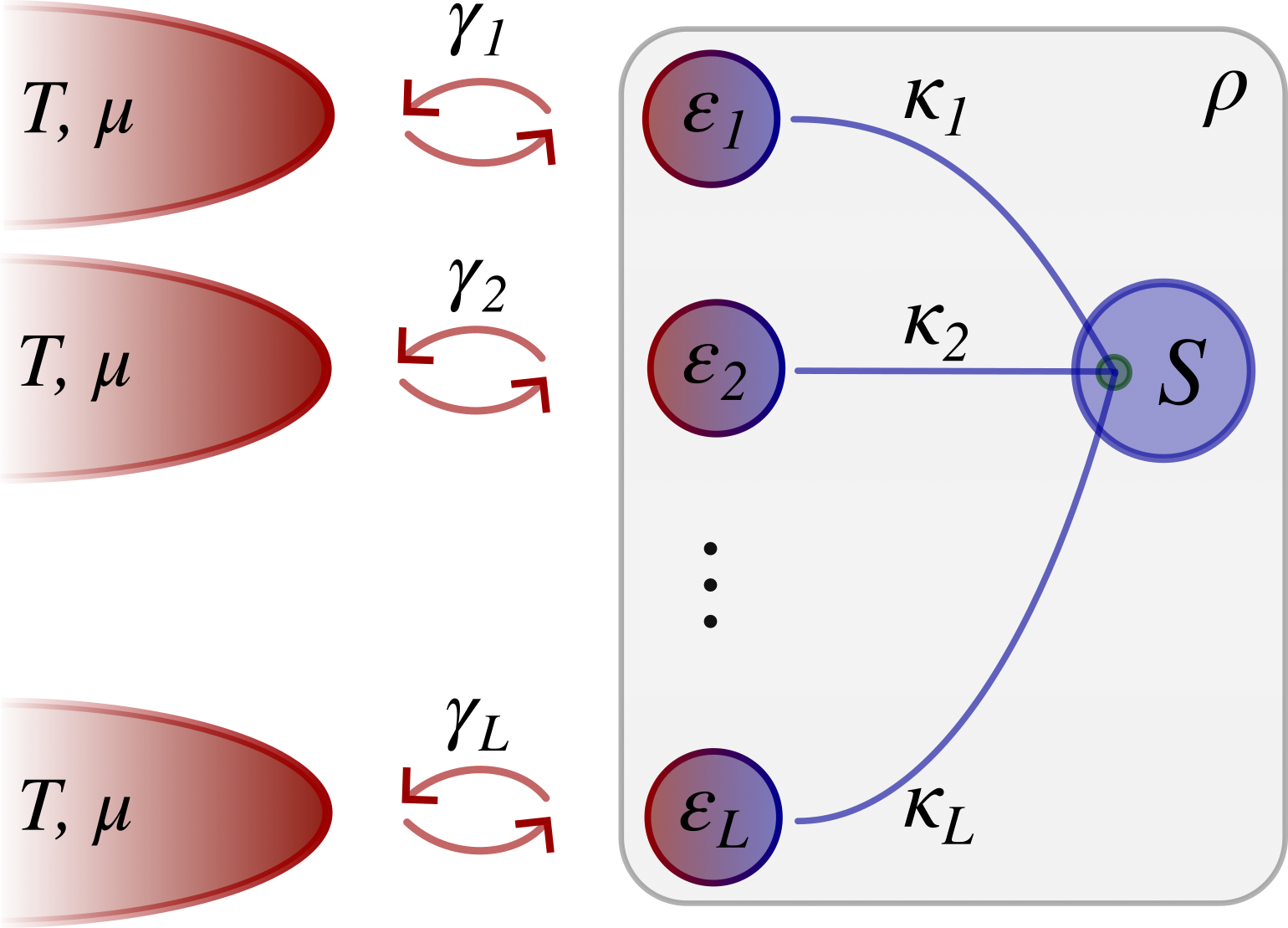}
  \caption{A central system $S$ coupled to a reservoir approximated in the mesoscopic-leads formalism. All $L$ lead modes with on-site energies $\epsilon_k$ couple to the same site in the central system, with hopping interaction of strength $\kappa_k$. The lead modes are damped through their coupling to a residual reservoir, with damping rate $\gamma_k$. The extended system, comprised of the central system and the lead modes, in state $\rho$ is shaded in grey.}
 \label{fig: mesoleads_schematic}
\end{center}
\end{figure}
We consider a fermionic central system $\mathrm{S}$ described by a set of $N_\mathrm{S}$ annihilation operators {$\lbrace d_j\rbrace_{j=1}^{N_\mathrm{S}}$}. The system, with Hamiltonian $H_\mathrm{S}(t)$, may be interacting and driven externally. The system is coupled locally to $N_\mathrm{R}$ thermal reservoirs, indexed by $\alpha$, with Hamiltonians (in natural units, $\hbar = 1$, $k_\mathrm{B} = 1$) $H_\mathrm{R_\alpha} = \sum_{n=1}^{\infty} \omega_{\alpha,n} b_{\alpha,n}^\dagger b_{\alpha,n}$, via the interaction $H_{\mathrm{SR}_\alpha}(t) = \sum_{n=1}^{\infty} \lambda_{\alpha,n}(t) d^\dagger_{p_\alpha} b_{\alpha,n} + \mathrm{h.c.}$, where  ${p_\alpha}$ denotes the index of the system site that the reservoir $\alpha$ couples to. The corresponding reservoir spectral density is given by $\mathcal{J}_\alpha(\omega) = 2\pi \sum_{n=1}^{\infty} |\lambda_{\alpha,n}|^2 \delta(\omega - \omega_{\alpha, n})$. 

In the mesoscopic-leads formulation each thermal reservoir $\alpha$, at temperature $T_\alpha$ and chemical potential $\mu_\alpha$, is modeled by a mesoscopic lead $\alpha$ with a finite number $l_\alpha$ of fermionic modes{, each with a residual reservoir acting on it.} The lead modes are described with annihilation operators $\lbrace a_{\alpha, k}\rbrace_{\alpha = 1, k=1}^{N_\mathrm{R}, l_\alpha}$ and have self-energies $\epsilon_{\alpha,k}$, so that $H_{\mathrm{L}_\alpha} = \sum_{k=1}^{l_\alpha} \epsilon_{\alpha,k} a_{\alpha, k}^\dagger a_{\alpha, k}$. The coupling between the system and each reservoir is replaced with the respective system-lead interaction $H_{\mathrm{SL}_\alpha}(t) = \sum_{k=1}^{l_\alpha} \kappa_{\alpha,k}(t) d^\dagger_{p_\alpha}a_{\alpha, k} + \mathrm{h.c.}$
Crucially, the residual reservoirs of each lead mode have a flat spectral density \cite{garraway_decay_1997, garraway_nonperturbative_1997} and effective damping rate $\gamma_{\alpha,k}$ (typically one sets $\gamma_{\alpha,k}= \epsilon_{\alpha, k+1} - \epsilon_{\alpha, k}$)~\cite{gruss_landauers_2016, elenewski_communication_2017, gruss_communication_2017, wojtowicz_open-system_2020, elenewski_performance_2021, wojtowicz_dual_2021}. Then, the coupling rate between the system site and modes in lead $\alpha$ is given by $\kappa_{\alpha,k} = \sqrt{\mathcal{J}_\alpha(\omega) {\gamma_{\alpha,k}}/2 \pi}$. 
Thus, each mode in the lead undergoes damping through interaction with a local Markovian reservoir, as illustrated schematically in Fig.~\ref{fig: mesoleads_schematic}.

For sufficiently large $l_\alpha$, $\gamma_{\alpha,k}$ becomes small, so that the residual reservoirs $(\mathrm{RR})$ may be traced out. The remaining extended system, described by the reduced density matrix $\rho$, is composed of the system modes as well as the lead modes. Its dynamics is generated by the Liouvillian superoperator $\mathcal{L}$, so that $\frac{\mathrm{d}\rho}{\mathrm{d}t}= \mathcal{L}\left[ \rho \right] $, in GKSL form 
\begin{equation}
\label{eq: ME_state}
    \begin{split}
    \mathcal{L}\left[ \rho \right] & = -i\left[H,\rho\right]+\sum_{\alpha=1}^{N_\mathrm{R}} \sum_{k=1}^{l_\alpha} \mathcal{L}_{\alpha, k }\left[ \rho \right],\\
     \mathcal{L}_{\alpha, k} \left[ \rho \right] &= 
     \sum_{\sigma 
     \in \lbrace+, -\rbrace} \mathcal{D}\left[{L_{\alpha, k}^\sigma} \right]\rho,
    \end{split}
\end{equation}
where $H(t) = H_\mathrm{S}(t) + \sum_{\alpha=1}^{N_\mathrm{R}} H_{\mathrm{SL}_\alpha}(t) + \sum_{\alpha=1}^{N_\mathrm{R}}H_{\mathrm{L}_\alpha}$, ${L_{\alpha, k}^+} = \sqrt{\gamma_{\alpha, k} f_{\alpha, k}} a_{\alpha, k}^ \dagger $ and ${L_{\alpha, k}^-} = \sqrt{\gamma_{\alpha, k} (1-f_{\alpha, k}}) a_{\alpha, k} $ with Fermi-Dirac occupation $f_{\alpha, k} = f_\alpha(\epsilon_k) = \left(e^{\left(\epsilon_{\alpha,k} -\mu_\alpha\right)/T_\alpha} + 1\right)^{-1}$. For any state $\rho$ and jump operator $L$ the superoperator $\mathcal{D}$ acts as $\mathcal{D}\left[L\right]\rho = L \rho L^ \dagger  - \frac{1}{2}\lbrace  L^ \dagger  L, \rho \rbrace$. 
 
Here it is important to highlight that, contrary to the system-lead dynamics, the lead-reservoir dynamics is incoherent. Typically, in GKSL master equations, there is an assumption of coherent evolution for the system, whereas the environment, represented by reservoirs, is assumed to be incoherent. However, if the interaction between the system and its environment is strong, then this distinction is no longer plausible, as the system and the environment hybridize. The coherent interaction between the lead modes and the system is thus a key feature of the mesoscopic-leads approach, enabling a natural modeling of systems in the strong-coupling regime.
That is, although the state of the extended system evolves under a GKSL ME, strong-coupling effects within the extended system are still captured in the dynamics of the lead modes and their respective coupling to the system, as in a so-called Markovian embedding \cite{woods_mappings_2014}. 

\subsection{State dynamics in non-interacting fermionic systems}
\label{sec: state_dynamics_gauss}
If the system $\mathrm{S}$ is non-interacting,  the Hamiltonian of the extended system may  be written in a quadratic form
\begin{equation}
    H=\sum_{i,j= 1}^{N_\mathrm{S}+L} {H}_{ij}c_i^ \dagger  c_j, 
\end{equation}
with annihilation operators describing the extended system $\lbrace c_i\rbrace_{i=1}^{N_\mathrm{S}+L}$ and $L = \sum_{\alpha = 1}^{N_\mathrm{R}} l_\alpha$, so that $\lbrace f_1, \dots, f_{N_\mathrm{S}}, a_{1,1}, \dots, a_{1,l_1}, \dots a_{N_\mathrm{R},1}, \dots, a_{N_\mathrm{R},l_{N_\mathrm{R}}}\rbrace \to \lbrace c_1, \dots, c_{N_\mathrm{S}+L} \rbrace $. 
Many quantities of interest, such as thermodynamic currents in the extended system, may then be described solely in terms of the covariance matrix $C$, with size $\left((N_\mathrm{S}+L)\times (N_\mathrm{S}+L)\right)$ and matrix elements $C_{ij} = \langle c_j^\dagger c_i \rangle_\rho$, rather than the full state  $\rho$, with size $(2^{N_\mathrm{S}+L}\times 2^{N_\mathrm{S}+L})$. Switching to the Heisenberg picture, the unconditional dynamics of the covariance matrix $C$ is governed by the ME
\begin{equation}
\label{eq: def_uc_C}
    \frac{\mathrm{d}C_{ij}}{\mathrm{d}t} =\left\langle\frac{\mathrm{d}c_j^ \dagger  c_i}{\mathrm{d}t}\right\rangle = \left\langle i\left[H,\rho\right]+ \sum_{\alpha = 1}^{N_\mathrm{R}}\sum_{k=1}^{l_\alpha}\mathcal{L}_{\alpha,k}^+\left[c_j^ \dagger  c_i\right]\right\rangle,
\end{equation}
where $\mathcal{L}_{\alpha,k}^+$ is the adjoint dissipator to $\mathcal{L}_{\alpha,k}$, satisfying $\mathrm{Tr}\left[A \mathcal{L}\left[ B\right]\right] = \mathrm{Tr}\left[\mathcal{L}^+\left[ A \right] B\right]$ for an arbitrary operator $A$, and $\mathcal{L}^+_\alpha \left[A\right]= \sum_{k=1}^{l_\alpha} \mathcal{L}_{\alpha,k}^+\left[A\right]$. In particular,
\begin{equation}
     \mathcal{L}^+_{\alpha, k }\left[A\right] =  \sum_{\sigma \in \lbrace +, - \rbrace}\mathcal{D}^+\left[L_{\alpha, k}^\sigma  \right]A,
\end{equation}
where $\mathcal{D}^+\left[L\right]A = L^\dagger A L - 
\frac{1}{2}\lbrace L^\dagger L, A \rbrace$.
One can readily show that the covariance matrix $C$ of the extended system evolves under the Lyapunov differential equation \cite{lyapunov_general_1994}
\begin{equation}
\label{eq: Lyapunov}
    \frac{\mathrm{d}C}{\mathrm{d}t} = -(WC + CW^ \dagger ) + F,
\end{equation}
where $W=iH+\frac{1}{2}\Gamma$, with diagonal matrices $\Gamma_{kk} = \gamma_k$ and $F_{kk} = \gamma_k f_k$. 
In the steady-state $\mathrm{d}C/\mathrm{d}t = 0$, so the covariance matrix solves the algebraic equation
\begin{equation}
    WC + C W^\dagger = F.
\end{equation}
 
\section{Quantum-jump trajectories in non-interacting fermionic systems}
\label{sec: cond_evo}
\begin{figure*}[t]
\begin{center}
\includegraphics[width=\linewidth]{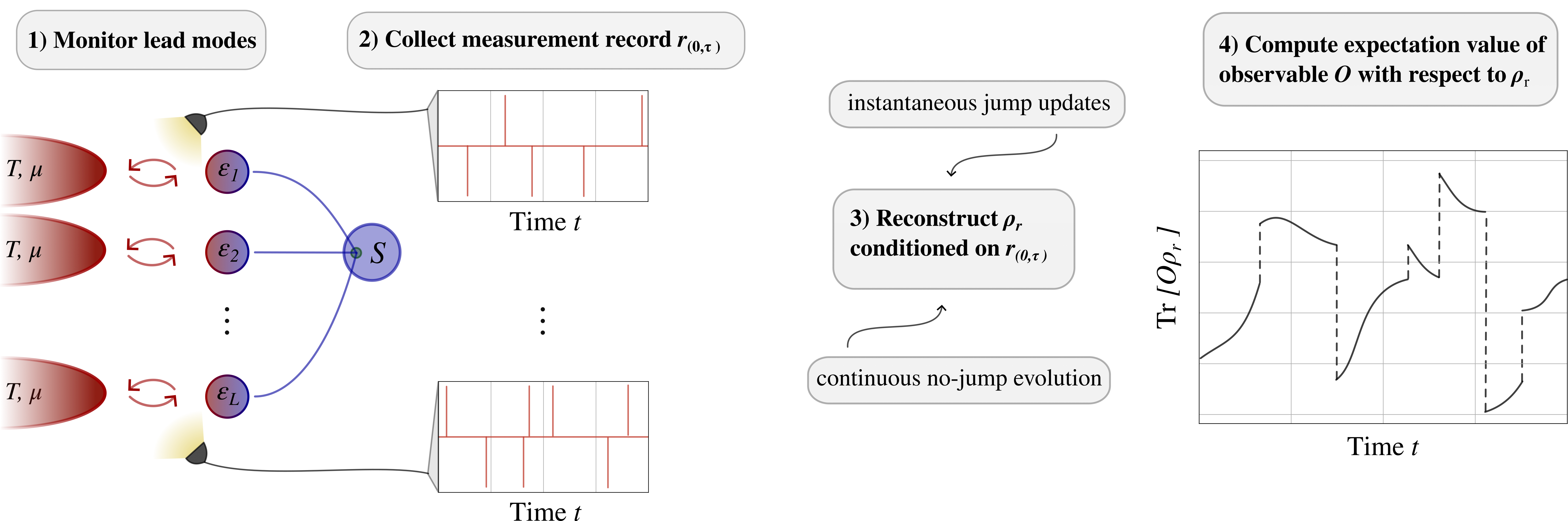}
  \caption{Particle transfer between each lead mode and its associated residual reservoir is monitored and recorded in the measurement record $r$. The expectation value of an observable $O$ is computed with respect to the state conditioned on $r$. When a quantum jump is recorded, the state undergoes an abrupt update, leading to an instantaneous change in the expectation value of the observable. During the intervals between jumps, the state evolves smoothly following the no-jump conditional evolution.}
 \label{fig: observable_trajectory}
\end{center}
\end{figure*}
For a quantum jump unraveling, it is convenient to split the Liouvillian $\mathcal{L}$ into a term representing quantum jumps $\mathcal{L}_1$, signaled by clicks in a classical detector, and into a term $\mathcal{L}_0$ describing a smooth no-jump evolution between two consecutive jumps, as indicated in Fig.~\ref{fig: observable_trajectory}, so that 
\begin{equation}
\label{eq: ME_state_split}
    \begin{split}
    \mathcal{L}\left[ \rho \right]&= \left(\mathcal{L}_0 + \mathcal{L}_1\right)\left[ \rho \right], \\
    \mathcal{L}_0 \left[ \rho \right] &= -i\left[H,\rho\right]-\frac{1}{2} \sum_{\alpha=1}^{N_\mathrm{R}} \sum_{k=1}^{l_\alpha} \sum_{\sigma 
     \in \lbrace+, -\rbrace}\left\lbrace {L_{\alpha,k}^\sigma}^\dagger L_{\alpha,k}^\sigma, \rho\right\rbrace, \\
    \mathcal{L}_1 \left[ \rho \right] &= \sum_{\alpha=1}^{N_\mathrm{R}} \sum_{k=1}^{l_\alpha} \sum_{\sigma 
     \in \lbrace+, -\rbrace} L_{\alpha,k}^\sigma \rho {L_{\alpha,k}^\sigma}^\dagger.
    \end{split}
\end{equation}
Then the solution to Eq.~\eqref{eq: ME_state} with initial state $\rho(t=0)= \rho_0$, generally given by $\rho(t) = e^{\mathcal{L}t}\rho_0$ for a time-independent $\mathcal{L}$, may be expanded in a Dyson series as
\begin{equation}
    \begin{split}
        \rho(t) &= e^{\mathcal{L}_0 t }\rho_0 + \int_0^t \mathrm{d}t_1 e^{\mathcal{L}_0 (t-t_1) } \mathcal{L}_1 e^{\mathcal{L}_0 t_1 }\rho_0\\
        &+\int_0^t \mathrm{d}t_2 \int_0^{t_2} \mathrm{d}t_1  e^{\mathcal{L}_0 (t-t_2) } \mathcal{L}_1 e^{\mathcal{L}_0 (t_2 - t_1) } \mathcal{L}_1 e^{\mathcal{L}_0 t_1 }\rho_0 + \dots,
    \end{split}
\end{equation}
which is the ensemble average over all possible trajectories with an increasing number of jumps in the interval $\left[0,t\right]$. The expansion illustrates why the above solution is typically denoted as the unconditional dynamics --- it is ignorant about whether any jumps occurred or not, and if they did, when they occurred, and thus may be interpreted as an unselective measurement.
We now consider single trajectories, for which the system's state evolves smoothly, occasionally interrupted by random quantum jumps. 
The jumps correspond to detection events in the environment, like the emission or absorption of particles from thermal reservoirs, as shown in Fig.~\ref{fig: observable_trajectory}.
Thus, quantum state trajectories correspond to the evolution of the state conditioned on a single measurement record.
Here, we examine a scenario involving the monitoring of particle exchange between all lead modes and their associated residual reservoirs.
\subsection{Gaussian initial states}
\label{sec: Gaussian_initial_states}
Crucially, we assume that the extended system is in a Gaussian fermionic state initially. The state then evolves along quantum jump trajectories, in which all processes involved are Gaussianity-preserving as shown in Appendix~\ref{sec:Gaussianity_appendix}. Therefore, at all times $t$, the density matrix may be expressed as
\begin{equation}
    \rho(t) = \frac{e^{-\mathbf{c}^ \dagger M(t)\mathbf{c}}}{Z(t)},
\end{equation}
where $\mathbf{c}$ is the vector with entries being the system-lead operators $\mathbf{c}_i = c_i$. The matrix $M(t)$ and the partition function $Z(t)$, ensuring normalization of the state, are given by
\begin{equation}
    M(t) = \log\left(\frac{1-C(t)}{C(t)}\right), \hspace{0.5cm} Z(t) = \frac{1}{\det\left[1 - C(t)\right]}.
\end{equation}
Since now the state $\rho(t)$ is fully determined by its covariance matrix $C(t)$, all formulas can only depend on matrices that are of size $\left((N_\mathrm{S}+L)\times (N_\mathrm{S}+L)\right)$, and thus they can be used efficiently, even for systems involving many modes.
Furthermore, we emphasize that the subsequent findings are applicable to the stochastic jump dynamics of any fermionic non-interacting system subjected to thermal boundary driving, provided that the initial state is Gaussian.

\subsection{Conditional evolution}
In the mesoscopic-leads formalism, quantum jump trajectories are obtained for the set of measurement operators
\begin{equation}
\label{eq: meas_ops}
    \begin{split}
        \Omega_0 &= \mathbb{1} - \frac{\mathrm{d}t}{2}\sum_{k = 1}^L  \sum_{ \sigma\in \lbrace +, - \rbrace}  {L^\sigma_k}^\dagger L^\sigma_k - i\mathrm{d}t H,\\
        \Omega_k^\sigma &= \sqrt{\mathrm{d}t} L^\sigma_k \hspace{1cm} \text{for $1 \leq k \leq L+1$},
    \end{split}
\end{equation}
where $ L^+_k = \sqrt{\gamma_k f_k}c_k^\dagger$ and $L^-_k = \sqrt{\gamma_k (1-f_k)} c_k$ are the jump operators associated with absorption/emission events respectively and $\mathbb{1}$ denotes the identity operator. Note that we have combined the lead index $\alpha$ with the index specifying the mode within the corresponding lead into a single index $k$, which runs over all modes across all leads.
The measurement record is denoted by $r_{(0,\tau)} = \lbrace(t_J, k, \sigma); 0 < t_J < \tau \rbrace$, recording a jump at time $t_J$ in lead mode $k$, where the jump signals particle transfer either on ($\sigma = +$) or off $(\sigma = -)$ it. Note that we neglect the probability of multiple jumps at one time.
Thus, at every time $t\in(0,\tau)$ our knowledge about the system's state must be updated: either conditioned on no-jump being observed or conditioned on a jump just having been registered. 

Under these measurement operators, the  evolution of $\rho$ unravels into a `no-jump' evolution with instantaneous jump-induced updates for the conditioned density matrix $\rho_r$ which in Lindblad-form (see Eq.~\eqref{eq: ME_state}) becomes the stochastic jump equation
\begin{equation}
\label{eq: rho_stochastic}
\begin{split}
    \mathrm{d}\rho_r =& -i\left[H, \rho_r\right]\mathrm{d}t-\frac{1}{2}\sum_{k=1}^L \sum_{ \sigma\in \lbrace +, - \rbrace} \mathcal{H}\left[{L^\sigma_k}^ \dagger  L^\sigma_k\right]\rho_r \mathrm{d}t\\
     &+ \sum_{k=1}^L \sum_{ \sigma\in \lbrace +, - \rbrace}  \mathcal{G}\left[L^\sigma_k\right]\rho_r \mathrm{d}N^\sigma_k.
\end{split}
\end{equation}
where the superoperators, $\mathcal{G}$ and $\mathcal{H}$, acting on a state $\rho_r$, are defined as
\begin{equation}
\label{eq: G_H_Schroedinger}
    \begin{split}
        \mathcal{G}[O]\rho_r&=\frac{O\rho_r O^ \dagger }{\mathrm{Tr}[O\rho_r O^ \dagger ]}-\rho_r\\
    \mathcal{H}[O]\rho_r&=O\rho_r + \rho_r O^ \dagger  - \mathrm{Tr}[O\rho_r + \rho_r O^ \dagger ]\rho_r.
    \end{split}
\end{equation}
{The superoperators $\mathcal{G}$ and $\mathcal{H}$ describe two types of measurement backaction: $\mathcal{G}$ corresponds to instantaneous state updates that occur when a jump is registered ($\mathrm{d}N_k^\sigma = 1$), while $\mathcal{H}$ represents the continuous, smooth effect between jumps ($\mathrm{d}N_k^\sigma = 0$), since the absence of jump detection still provides information about the system.}
The stochastic increments of the point processes $\mathrm{d}N^\sigma_k $ obey \cite{wiseman_quantum_2009}
\begin{align}
    \mathrm{E}\left[\mathrm{d}N\right] &= \langle{\Omega_k^\sigma}^\dagger \Omega_k^\sigma\rangle_\rho,\\
     \mathrm{d}N^{\sigma^\prime}_{k^\prime}  \mathrm{d}N^\sigma_k &=  \mathrm{d}N^\sigma_k \delta_{\sigma\sigma^\prime}\delta_{kk^\prime},
\end{align}
where $\mathrm{E}$ denotes a classical expectation value, i.e.~an average over random realizations of the measurement record, corresponding to the expectation value computed with respect to the ensemble average state $\rho$.
Here and in the following, we assume perfect detection efficiencies in all measurement channels.  A general treatment accounting for imperfect detection or different measurement set-ups is provided in Appendix~\ref{sec: imperfect_detection}.
The expectation values $\langle\cdot\rangle_r$, conditioned on the measurement record, are connected to the averages of the corresponding measurement operators $\langle\cdot\rangle_r = \mathrm{Tr}\left[\rho_r \cdot\right]$ (Eq.~\eqref{eq: meas_ops}). For a Gaussian fermionic system, both ensemble averages and individual trajectories can be characterized using the covariance matrix, as will become apparent in the following sections. Specifically, for single trajectories, this matrix is computed concerning the conditioned state, denoted as $C^r_{ij} =\mathrm{Tr}\left[c_j^\dagger c_i\rho_r\right]$. We will subsequently discuss the evolution of $C^r$ between jumps and its update when jumps are recorded.

\subsubsection{No jump-conditioned evolution}
\label{sec: no-jump}
We will now describe how the covariance matrix $C^r$ of the extended system, conditioned on the measurement record $r$, evolves in the time interval between two recorded jump events, $(t_{J_{m-1}}, k_{m-1}, \sigma_{m-1})$ and $(t_{J_m}, k_m, \sigma_m)$. 
Note, that the conditional no-jump evolution in Gaussian fermionic systems has been studied also in \cite{coppola_conditional_2023}. For $t\in (t_{J_{m-1}}, t_{J_m})$, $\mathrm{d}N^\sigma_k(t)=0$ $\forall k, \sigma$, so that $C^r$ evolves according to 
\begin{equation}
\label{eq: C_nojump_app}
\begin{split}
    \mathrm{d}C^r_{ij}(t)& 
    =  i\left\langle\left[H, c_j^\dagger c_i \right]\right\rangle_r \mathrm{d}t\\
    &-\frac{1}{2}\sum_{k=1}^L \sum_{ \sigma\in \lbrace +, - \rbrace}  \left\langle\mathcal{H}^+\left[\rho_r,{L^\sigma_k}^\dagger L^\sigma_k\right]c^ \dagger _j c_i \right\rangle_r\mathrm{d}t,
\end{split}
\end{equation}
where the adjoint superoperator to $H$ (see Eq.~\eqref{eq: G_H_Schroedinger}), $\mathcal{H}^+$, acting on any operator $A$ is defined as
\begin{equation}
    \begin{split}
    \mathcal{H}^+[\rho_r,O]A&=OA + A O^ \dagger  - \mathrm{Tr}\left[O\rho_r + \rho_r O^ \dagger \right]A.
    \end{split}
\end{equation}
Expanding Eq.~\eqref{eq: C_nojump_app}, one obtains terms of fourth order in the annihilation and creation operators of the extended system. 
Crucially, we now make use of the fact that the initial state is assumed to be Gaussian and that all operations along a single trajectory preserve the state's Gaussianity~\cite{bravyi_lagrangian_2005}. 
Note, that we can reduce fourth order correlators with respect to a Gaussian state $\phi$ to second order or lower, by using Wick’s theorem \cite{wick_evaluation_1950}: Let $O_i$ denote operators which are arbitrary linear combinations of bosonic or fermionic creation and annihilation operators. Wick’s theorem states that
\begin{equation}
\begin{split}
    \langle  O_1 O_2 &O_3 O_4\rangle_\phi =   \langle  O_1 O_2\rangle_\phi\langle  O_3 O_4\rangle_\phi \pm \langle  O_1 O_3\rangle_\phi\langle  O_2 O_4\rangle_\phi \\
    &+\langle  O_1 O_4\rangle_\phi\langle  O_2 O_3\rangle_\phi -2\langle  O_1\rangle_\phi\langle  O_2\rangle\langle  O_3\rangle_\phi\langle  O_4\rangle_\phi,
\end{split}
\end{equation}
where the the upper (lower) sign is for bosons (fermions).
We find a Riccati-type differential equation for the no-jump conditioned covariance matrix of the extended system
\begin{equation}
\begin{split}
    \frac{\mathrm{d}C^r}{\mathrm{d}t} =& -(VC^r + C^r V^ \dagger ) + C^r B C^r,
\end{split}
\end{equation}
where $B = \Gamma - 2F$, $V = i H + \frac{1}{2}B = W-F$. 
In between jumps, the survival probability $p$---the probability for no jump to occur up to time $t$---evolves under the differential equation
\begin{equation} 
\begin{split}
    \frac{\mathrm{d} p}{\mathrm{d}t} &= - p \sum_{k=1}^L \sum_{ \sigma\in \lbrace +, - \rbrace}  \left\langle {L^\sigma_k}^\dagger {L^\sigma_k} \right\rangle_r, \\
    \end{split}
 \end{equation} 
 where we have used that $p$ is given by the trace of the unnormalized density matrix between jumps.
Therefore, the survival probability decays as
\begin{equation}
\begin{split}
    p(t) &= p(t_{J_{m-1}})  \exp\left(-\int_{t_{J_{m-1}}}^{t} K(s)\mathrm{d} s\right),
\end{split}
\end{equation}
where
\begin{equation}
\begin{split}
    K(s) &=   \mathrm{Tr}\left[F\bar{C}^r(s)\right] + \mathrm{Tr}\left[(\Gamma-F)C^r(s)\right],
\end{split}
\end{equation}
and $\bar{C}^r= 1-C^r$.
\subsubsection{Jump-conditioned updates}
\label{sec: jump2}
Upon recording a jump of type $\sigma_{m}$ in the lead mode $k_m$ at time $t_{J_m}$, stored in the measurement record as $(t_{J_m}, k_{m}, \sigma_{m})$, $C^r$ is updated instantaneously. At $t = t_{J_m}$, $\mathrm{d}N^\sigma_{k}(t_{J_m})=\delta_{k k_{m}}\delta_{\sigma \sigma_{m}}\in \lbrace 0,1 \rbrace$ and therefore
\begin{equation}
\begin{split}
    \mathrm{d}C^r_{ij}(t_{J_m}) 
     =& \left\langle \mathcal{G}^+\left[\rho_r,{L^{\sigma_{m}}_{k_{m}}}\right]c^ \dagger _j c_i\right\rangle_r(t_{J_m}) .
\end{split}
\end{equation}
where the adjoint superoperator to $G$ (see Eq.~\eqref{eq: G_H_Schroedinger}), $\mathcal{G}^+$, acting on any operator $A$ is defined as
\begin{equation}
    \begin{split}
    \mathcal{G}^+[\rho_r,O]A&=\frac{O^ \dagger  A O}{\mathrm{Tr}\left[O\rho_r O^ \dagger \right]}-A.
    \end{split}
\end{equation}
The update entails that the survival probability is instantaneously reset to 1 ($p(t^+_{J_m}) = p(t^-_{J_m}) + (1- p(t^-_{J_m})) = 1$), and the updated covariance matrix $C^r(t^+_{J_m}) = C^r(t^-_{J_m}) + \mathrm{d}C^r_{ij}(t_{J_m})$, where for any time-dependent quantity $A(t)$, we use the short-hand $A(t^{\pm}) = \lim_{s\to t^\pm} A(s)$.
If a jump onto the lead mode $k_{m}$ from its residual reservoir is recorded, so that  $\sigma_{m} = +$, then
\begin{equation}
    \label{eq: jump_update_C_+}
    \mathrm{d}C^r(t_{J_m}) =  \bar{C}^r \left(\frac{\mathrm{d}N^{+}(t_{J_m})}{\bar{C}^r}\right) \bar{C}^r.
\end{equation}
Otherwise, if one records a jump off the lead mode $k_{m}$ to its residual reservoir, so that  $\sigma_{m} = -$, then
\begin{equation}
      \mathrm{d}C^r(t_{J_m}) = - C^r \left(\frac{\mathrm{d}N^{-}(t_{J_m})}{C^r}\right) C^r.
\end{equation}
Above, we use the shorthand
\begin{equation}
    \begin{split}
        \left(\frac{\mathrm{d}N^{+}(t)}{\bar{C}^r} \right)_{kk} &= \frac{\mathrm{d}N^+{k}(t)}{1-C^r_{kk}} 
        ,\\
\left(\frac{\mathrm{d}N^{-}(t)}{C^r}\right)_{kk} &=
    \frac{\mathrm{d}N^-_{k}(t)}{C^r_{kk}}
    .
    \end{split}
\end{equation}
\subsubsection{Stochastic master equation for the covariance matrix}
Finally, we combine the results presented in Sec.~\ref{sec: no-jump} and Sec.~\ref{sec: jump2} to obtain the stochastic ME governing the trajectory of $C^r$
\begin{equation}
    \begin{split}
        \mathrm{d}C^r(t) = & \left[-(VC^r + C^r V^ \dagger ) + C^r B C^r\right]\mathrm{d}t\\
        &  +\bar{C}^r \left(\frac{\mathrm{d}N^{+}(t)}{\bar{C}^r}\right) \bar{C}^r \\
        & - C^r \left(\frac{\mathrm{d}N^{-}(t)}{C^r}\right) C^r.
    \end{split}
\end{equation}
Using It\^o's Lemma, by which $\mathrm{E}\Big[ dN^\sigma_k(t) f(\rho_r(t))\Big] = \mathrm{d}t \mathrm{E}\Big[ \langle {L^\sigma_k}^\dagger L^\sigma_k \rangle_r(t) f(\rho_r(t))\Big]$, one recovers Eq.~\eqref{eq: Lyapunov} for the unconditional evolution of $C$.
In Appendix~\ref{sec: monte_carlo_solver}, a short description is provided for the numerical implementation of trajectory sampling relying on the covariance matrix.

\section{Bayesian estimation of thermodynamic currents}
\label{sec: cond_thermo}

In the presence of time-dependent driving or multiple reservoirs at different temperatures or chemical potentials, currents of energy and particles will flow through the system, irreversibly producing entropy. The average values of these currents are given by appropriate expectation values with respect to the unconditional density matrix. Importantly, the mesoscopic-leads formalism can reproduce exact NEGF results for these average currents so long as the number of lead modes $L$ is large enough, as shown in previous works \cite{brenes_tensor-network_2020, lacerda_quantum_2022, brenes_particle_2023}. 
{Note that the currents of interest in our framework are those exchanged between the lead modes and the residual reservoir. This distinction is important, as we argue that the currents measured in mesoscopic experiments, e.g.~by an ammeter, correspond more closely to these internal reservoir currents---especially in the strong-coupling regime. In this regime, the presence of coherences and hybridization blurs the boundary between the system and its environment, making a clear separation between them effectively impossible. In this work, we aim to capture the fluctuations of the time-averaged current that would be measured in an experimental setting. These experimentally accessible fluctuations can be obtained from our theoretical framework by analyzing the statistics of the integrated charge along quantum trajectories. }

{In this section, we evaluate fluctuations of thermodynamic currents along individual trajectories using the mesoscopic-leads formalism. We interpret the conditional state $\rho_r$ as the best estimate of the system's state, given the measurement record. The current is computed with respect to this conditional state, which evolves as new data is acquired, reflecting a Bayesian update of the observer’s knowledge, as commonly understood in the theory of quantum trajectories~\cite{wiseman_quantum_2009}.} As we will show, the resulting stochastic currents reduce to standard results on average, and yield thermodynamically consistent results for the fluctuations (see Sec.~\ref{sec: EP_TPM}). To this end, we detail how to express these currents in relation to the trajectory covariance matrix $C^r$. 
In the main text, we present the results assuming perfect detection in all measurement channels. More comprehensive findings considering imperfect detection and different measurement configurations can be found in Appendix~\ref{sec: imperfect_detection}.

It is important to point out that our focus here is on the currents exchanged between the extended system and the residual reservoirs, termed external currents, as opposed to the internal currents exchanged between the system and the leads. We argue that it is the external currents that correspond to those detectable in experiments, especially in the strong coupling regime, where the system hybridises with its environment. 
For an in-depth discussion of internal and external currents in the mesoscopic-leads formalism, please refer to Refs.~\cite{lacerda_quantum_2022,lacerda_entropy_2023}.
\subsubsection{Particle current}
\label{sec: cond_JP}
The average (unconditional) particle current $I_N(t)$ flowing into the extended system is defined through the continuity equation for the total number operator $N = \sum_{j=1}^{N_\mathrm{S}+L} c_j^ \dagger  c_j$
\begin{equation}
\label{eq: JN_uc}
\begin{split}
    \frac{\mathrm{d}\langle N \rangle}{\mathrm{d}t} = \sum_{\alpha = 1}^{N_\mathrm{R}} I_{N_\alpha}(t)\,,
\end{split}
\end{equation}
where $I_{N_\alpha}(t) = \langle \mathcal{L}_\alpha^+\left[ N\right]\rangle$ denotes the particle current into lead $\alpha$ and $ I_{N}(t) = \sum_{\alpha = 1}^{N_\mathrm{R}} I_{N_\alpha}(t)$ is the net particle current. Note that since $\left[H, N\right] = 0$, $\mathrm{Tr}\left[N\mathcal{L}_0 \left[ \rho(t)\right]\right] = 0$. This can be understood intuitively, as hopping interactions within the extended system conserve the total particle number. Therefore, the total particle current is given soley as the sum over the net particle currents flowing into the extended system from the residual reservoirs.
Similarly, defined through a continuity equation, the net particle current conditioned on the measurement record and thus along a single trajectory, $I^{r}_N(t)$, is given by
\begin{equation}
\label{eq: JN_c}
\begin{split}
    \frac{\mathrm{d}\langle N \rangle_r}{\mathrm{d}t} = \sum_{\alpha = 1}^{N_\mathrm{R}} I^{r}_{N_\alpha}(t)\,,
\end{split}
\end{equation}
so that $I^{r}_N(t) = \sum_{\alpha = 1}^{N_\mathrm{R}} I^{r}_{N_\alpha}(t)$, where
\begin{equation}
\begin{split}
    I^{r}_{N_\alpha}(t) \mathrm{d}t=&   -\frac{1}{2}\sum_{k=1}^{l_\alpha} \sum_{ \sigma\in \lbrace +, - \rbrace} \mathrm{Tr}\left[N\mathcal{H}\left[{L^\sigma_k}^ \dagger  L^\sigma_k\right]\rho_r \mathrm{d}t
    \right]\\
     &+\sum_{k=1}^{l_\alpha} \sum_{ \sigma\in \lbrace +, - \rbrace}  \mathrm{Tr}\left[N\mathcal{G}\left[L^\sigma_k\right]\rho_r \mathrm{d}N^\sigma_k \right]
\end{split}
\end{equation}
is the inferred particle current into lead $\alpha$.
We find 
\begin{equation}\label{eq_bayesian_current}
\begin{split}
     I^{r}_{N_\alpha}(t) \mathrm{dt} = & \mathrm{Tr}\left[C^r B_{\alpha}C^r - B_{\alpha}C^r\right] \mathrm{d}t \\
    &+  \mathrm{Tr}\left[ \bar{C}^r\left(\frac{\mathrm{d}N^+_{ {\alpha}}(t)}{\bar{C}^r}\right) \bar{C}^r \right] \\
    &-  \mathrm{Tr}\left[C^r\left(\frac{\mathrm{d}N^-_{{\alpha}}(t)}{C^r}\right)C^r\right],
\end{split}
\end{equation}
where $B_\alpha=\Gamma_\alpha - 2F_\alpha$ with
\begin{equation}
\begin{split}
    \left(\Gamma_\alpha \right)_{kk}, \left(F_\alpha \right)_{kk}= \begin{cases} 
      \Gamma_{kk}, F_{kk} &  \text{if $k$-mode in lead $\alpha$,} \\
      0 & \text{otherwise.}\\
   \end{cases}
\end{split}
\end{equation}
We use the shorthand
\begin{equation}
    \begin{split}
        \left(\frac{\mathrm{d}N^+_{ \alpha}(t)}{\bar{C}^r} \right)_{kk} &= \begin{cases}
    \frac{\mathrm{d}N^+_{k, \alpha}(t)}{\bar{C}^r_{kk}} & \text{ if $k$-mode in lead $\alpha$,} \\
    0 & \text{ otherwise.}
\end{cases}\\
\left(\frac{\mathrm{d}N^-_{ \alpha}(t)}{C}\right)_{kk} &= \begin{cases}
    \frac{\mathrm{d}N^-_{k, \alpha}(t)}{C^r_{kk}} & \text{ if $k$-mode in lead $\alpha$,} \\
    0 & \text{ otherwise.}
\end{cases}
    \end{split}
\end{equation}
The Bayesian nature of the current is apparent in Eq.~\eqref{eq_bayesian_current} by the fact that even the extraction of an electron (last term) does not necessarily imply a current of $-1$. 
This is because the covariance matrix also encompasses our uncertainty about the number of particles within the extended system prior to a detection event. Thus, even though a quantum jump definitely corresponds to the detection of an electron, this does not imply that the change in the occupation number of the extended system is $-1$, since initially there was some uncertainty as to the number of electrons in there.

Performing an ensemble average and making use of It\^o's lemma we find that the average (unconditional) particle current is given by
\begin{equation}
\begin{split}
    I_{N_\alpha}(t) = \mathrm{Tr}\left[F_\alpha - \Gamma_\alpha C(t)\right].
\end{split}
\end{equation}
When plugging in the explicit form of $F_\alpha$ and $\Gamma_\alpha$, we recover
\begin{equation}
\begin{split}
    I_{N_\alpha}(t) 
    = \sum_{k=1}^{l_\alpha} \gamma_{k, \alpha}  \langle  f_{k, \alpha} - c^ \dagger _{k, \alpha} c_{k, \alpha}\rangle,
\end{split}
\end{equation}
which is consistent with the standard unconditional continuity equation for the particle current, stated in Eq.~\eqref{eq: JN_uc}, and is in agreement with results derived in \cite{brenes_tensor-network_2020}.

\subsubsection{Energy current}
\label{sec: cond_JE}
The average (unconditional) total energy current $I_E(t)$ flowing into the extended system is defined through the continuity equation for the Hamiltonian of the extended system $H$
\begin{equation}
\label{eq: JE_uc}
\begin{split}
    \frac{\mathrm{d}\langle H \rangle}{\mathrm{d}t}
    & = \sum_{\alpha=1}^{N_\mathrm{R}}  I_{E_\alpha}(t),
\end{split}
\end{equation}
where $I_{E_\alpha}(t) = \langle \mathcal{L}_\alpha^+\left[ H\right]\rangle$ denotes the energy current into lead $\alpha$ and $I_{E}(t) = \sum_{\alpha=1}^{N_\mathrm{R}} I_{E_\alpha}(t)$ is the net energy current.
We find that the energy current into the extended system via lead $\alpha$ along a single trajectory, inferred on the basis of the measurement record as discussed in Sec.~\ref{sec: cond_JP}, is given by
\begin{equation}
\begin{split}
     I^{r}_{E_\alpha}(t) \mathrm{d}t = & \ \mathrm{Tr}\left[B_\alpha\left(C^r HC^r - \frac{1}{2}(HC^r+C^r H)\right)\right] \mathrm{d}t \\
    &+  \mathrm{Tr}\left[ H \bar{C}^r\left(\frac{\mathrm{d}N^+_\alpha(t)}{\bar{C}}\right)\bar{C}^r\right] \\
    &-\mathrm{Tr}\left[HC^r\left(\frac{\mathrm{d}N^-_\alpha(t)}{C^r}\right)C^r\right].
\end{split}
\end{equation}
The total inferred energy current is given by
\begin{equation}
\begin{split}
     I^{r}_E(t) = &\sum_{\alpha = 1}^{N_\mathrm{R}} I^{r}_{E_\alpha}(t). 
\end{split}
\end{equation}
Therefore the energy change along a single trajectory in time interval $\left[0,\tau\right]$ conditioned on the measurement record is given by 
\begin{equation}
\label{eq: energy_balance}
    \Delta E^r (\tau, 0) = \int_0^\tau  I^{r}_{E_\alpha}(t)\mathrm{d}t.
\end{equation}
Performing an ensemble average and using It\^o's lemma, we find that the unconditional energy current is given by
\begin{equation}
\begin{split}
     I_E(t) 
     &= \mathrm{Tr}\left[FH(t) - \frac{1}{2} \Gamma \left(C(t)H(t)+H(t)C(t)\right)\right].
\end{split}
\end{equation}
When plugging in the explicit form of the Hamiltonian, we recover
\begin{equation}
\label{eq: JE_unc}
\begin{split}
    I_E(t) =& \sum_{k=1}^L \gamma_{k} \epsilon_{k} \langle  f_{k} - c^ \dagger _{k} c_{k}\rangle\\
        &-\frac{1}{2}\sum_{k=1}^L \gamma_{k} \langle  \kappa_{kp}c^\dagger_p a_{k} + \kappa^*_{kp}a_{kp}^ \dagger  c_p\rangle,
        \end{split}
\end{equation}
in agreement with results derived in Ref. \cite{brenes_tensor-network_2020}.

\subsubsection{Heat current and measurement energy}
\label{sec: meas_work}
The heat current is typically defined as 
\begin{equation}
    \begin{split}
        I^Q(t) 
        &=
        \sum_{\alpha=1}^{N_\mathrm{R}} I^E_\alpha(t)- \mu_\alpha I^N_\alpha(t).
    \end{split}
\end{equation}
However, in the energy current balance along trajectories described by stochastic MEs involving jump operators that are not eigenoperators of the Hamiltonian, a specific term can be singled out \cite{horowitz_quantum-trajectory_2012, manzano_nonequilibrium_2015,levy_local_2014} that poses ambiguity in its interpretation—whether it should be considered as ``quantum heat'' current \cite{elouard_role_2017} or ``measurement work'' current\cite{horowitz_quantum-trajectory_2012, manzano_nonequilibrium_2015, manzano_quantum_2022} is not definitively established. 
It is given by \cite{manzano_quantum_2022}
\begin{widetext}
\begin{equation}
\label{eq: I_EM}
\begin{split}
    I_{E_\mathrm{M}}\mathrm{d}t = &-\frac{\mathrm{d}t}{2} \sum_{k=1}^L \sum_{\sigma \in \lbrace +, - \rbrace} \mathrm{Tr}\left[H\lbrace {L^\sigma_k}^\dagger L^\sigma_k, \rho_r\rbrace\right] 
    +\mathrm{d}t\sum_{k=1}^L \sum_{\sigma \in \lbrace +, - \rbrace}  \langle H \rangle_r \langle  {L^\sigma_k}^ \dagger  L^\sigma_k\rangle_r
    +\sum_{k=1}^L \sum_{\sigma \in \lbrace +, - \rbrace}  \mathrm{d}N^\sigma_k \frac{\mathrm{Tr}\left[{L^\sigma_k}^\dagger H_\mathrm{int} L^\sigma_k \rho_r\right]}{\langle  {L^\sigma_k}^ \dagger  L^\sigma_k\rangle_r} 
     \\
    &
    +\frac{1}{2}\sum_{k=1}^L \sum_{\sigma \in \lbrace +, - \rbrace} \mathrm{d}N^\sigma_k\frac{\mathrm{Tr}\left[\lbrace  {L^\sigma_k}^ \dagger  L^\sigma_k, H_0 \rbrace \rho_r\right]}{\langle  {L^\sigma_k}^ \dagger  L^\sigma_k\rangle_r} 
    -\sum_{k=1}^L \sum_{\sigma \in \lbrace +, - \rbrace} \mathrm{d}N^\sigma_k \langle  H\rangle_r,
\end{split}
\end{equation}
\end{widetext}
and is a result of the non-local nature of energy in interactions:
Here, the Hamiltonian of the extended system is split into $H= H_0  + H_\mathrm{int}$, where $H_0  = H_\mathrm{S} + H_\mathrm{L}$ is the bare Hamiltonian consisting of the Hamiltonian of the system and that of the leads, and $H_\mathrm{int} = H_\mathrm{SL}$ is the interaction between them.
This is important for the computation of the entropy production, which we explore in detail in Sec.~\ref{sec: EP_TPM}. There, one considers energy increments $\Delta E_k$ exchanged between system and environment, which are defined through 
\begin{equation}
    \left[H_0 , L_k\right] = -\Delta E_k L_k.
\end{equation}
If the chemical potential is nonzero, additionally, particle exchange between the reservoirs and the lead modes contributes to entropy production and
\begin{equation}
    \left[N,L_k\right] = - \Delta N_k L_k.
\end{equation}
Here, as in many other cases of physical interest, the set of jump operators is self-adjoint, and the jumps obey the local detailed balance condition for the corresponding pairs of operators, so that $L_k^+ = {L_k^-}^\dagger e^{\Delta s_k/2}$. Therefore, the entropy produced upon a jump in channel $k$ is given by
\begin{equation}
\begin{split}
     \Delta s_k 
    &= -\left(\frac{\Delta E_k}{T_r}- \mu_r \frac{\Delta N_k}{T_r}\right),
\end{split}
\end{equation}
where $T_r$ and $\mu_r$ are the temperature and chemical potential, respectively, of the residual reservoir that mode $k$ interacts with.  
For the conditional measurement energy current for lead $\alpha$ we find
\begin{widetext}
    \begin{equation}
\begin{split}
    I_{E_{\mathrm{M}, \alpha}}\mathrm{d}t
    = & \mathrm{Tr}\left[B_\alpha\left(C^r HC^r - \frac{1}{2}(HC^r+C^r H)\right)\right] \mathrm{d}t \\
    &+\mathrm{Tr}\left[\left(\frac{\mathrm{d}N^+_{\alpha}}{\bar{C}^r}\right)\bar{C}^rH_\mathrm{int} \bar{C}^r\right]
    -\mathrm{Tr}\left[\left(\frac{\mathrm{d}N^-_\alpha}{C^r}\right)C^r H_\mathrm{int} C^r\right]\\
    & -\frac{1}{2}\mathrm{Tr}\left[\left(\frac{\Lambda^+\mathrm{d}N^+_\alpha}{\bar{C}^r} \right)\left(\bar{C}^r H_0 C^r + C^r H_0 \bar{C}^r\right)\right]
     -\frac{1}{2}\mathrm{Tr}\left[\left(\frac{\mathrm{d}N^-_\alpha}{C^r}\right)\left(\bar{C}^r H_0 C^r + C^r H_0 \bar{C}^r\right)\right].
\end{split}
\end{equation}
\end{widetext}
The total measurement energy current is obtained as the sum over the contributions from the different leads 
\begin{equation}
\begin{split}
    I_{E_\mathrm{M}} = &\sum_{\alpha = 1}^{N_\mathrm{R}}I_{E_{\mathrm{M}, \alpha}}.
\end{split}
\end{equation}
Taking the ensemble average, using It\^o's lemma, we find
\begin{equation}
    E\left[I_{E_\mathrm{M}}\right] = - \frac{1}{2} \mathrm{Tr}\left[\Gamma(H_\mathrm{int}C+H_\mathrm{int}C)\right].
\end{equation}
Interestingly, when plugging in the explicit form of the system-lead interaction $H_\mathrm{int} = H_{\mathrm{SL}}(t) = \sum_{\alpha=1}^{N_\mathrm{R}} \sum_{k=1}^{l_\alpha} \kappa_{\alpha,k} d^\dagger_{p_\alpha} a_k + \mathrm{h.c.}$, as defined in Sec.~\ref{sec: mesoleads_formulation}, then we find
\begin{equation}
\label{average_measurement_energy}
    E\left[I_{E_\mathrm{M}}\right] =-\frac{1}{2}
    \sum_{\alpha=1}^{N_\mathrm{R}} \sum_{k=1}^{l_\alpha} \gamma_{\alpha,k} \langle \kappa_{\alpha,k} d^\dagger_{p_\alpha} a_k + \kappa_{\alpha,k}^* a_k^\dagger f_{p_\alpha} \rangle.
\end{equation}
$E\left[I_{E_\mathrm{M}}\right]$ is thus exactly the term appearing in the energy current due the coherent hopping interaction between the lead modes and their respective coupling site in the central system, see Eq.~\eqref{eq: JE_unc}. Indeed, this kind of term has also appeared in many other contexts, like the repeated interactions framework \cite{chiara_reconciliation_2018}.
\section{Entropy production along a trajectory}
\label{sec: EP_TPM}

By the second law of thermodynamics, changes in the entropy of the universe (i.e. a closed system) due to an irreversible process are positive. This, however, only holds on average and for meso- and microscopic set-ups, fluctuations around the average play an important role. In our setting, where currents exchanged between the extended system and its environment, one is bound to compute a stochastic entropy production which becomes a measure of irreversibility along single trajectories.

In general, the stochastic entropy production in the extended system along a single trajectory from time 0 to $\tau$ is given by
\begin{equation}
\label{eq: S_tot}
    S_\mathrm{tot}({r_{\left[0, \tau\right]}}) = \log\left(\frac{P_F(r)}{P_B(\Tilde{r})}\right),
\end{equation}
which depends on the log-ratio of probabilities for the forward trajectory $r_{\left[0, \tau\right]}$ and backwards trajectory $\Tilde{r}_{\left[0, \tau\right]}$ to occur (these trajectories are precisely defined below). By averaging the exponentiated negative of the total stochastic entropy production over the forward trajectories $r$, it is easy to show that
\begin{equation}
\label{eq: integral_FT}
    \langle  e^{-S_\mathrm{tot}({r_{\left[0, \tau\right]}})}\rangle_r = 1,
\end{equation}
a central result of stochastic thermodynamics~\cite{evans_fluctuation_2002, jarzynski_equalities_2011, ciliberto_experiments_2017, esposito_nonequilibrium_2009, campisi_fluctuation_2014}.
This is known as the integral fluctuation relation and in turn, by Jensen’s inequality $\langle e^x \rangle \geq e^{\langle x\rangle}$, leads to the second law inequality 
\begin{equation}
    \langle S_\mathrm{tot}({r_{\left[0, \tau\right]}}) \rangle_r \geq 0.
\end{equation}

\subsection{Two-point measurement scheme}

\begin{figure}[t]
\begin{center}
\includegraphics[width=\linewidth]{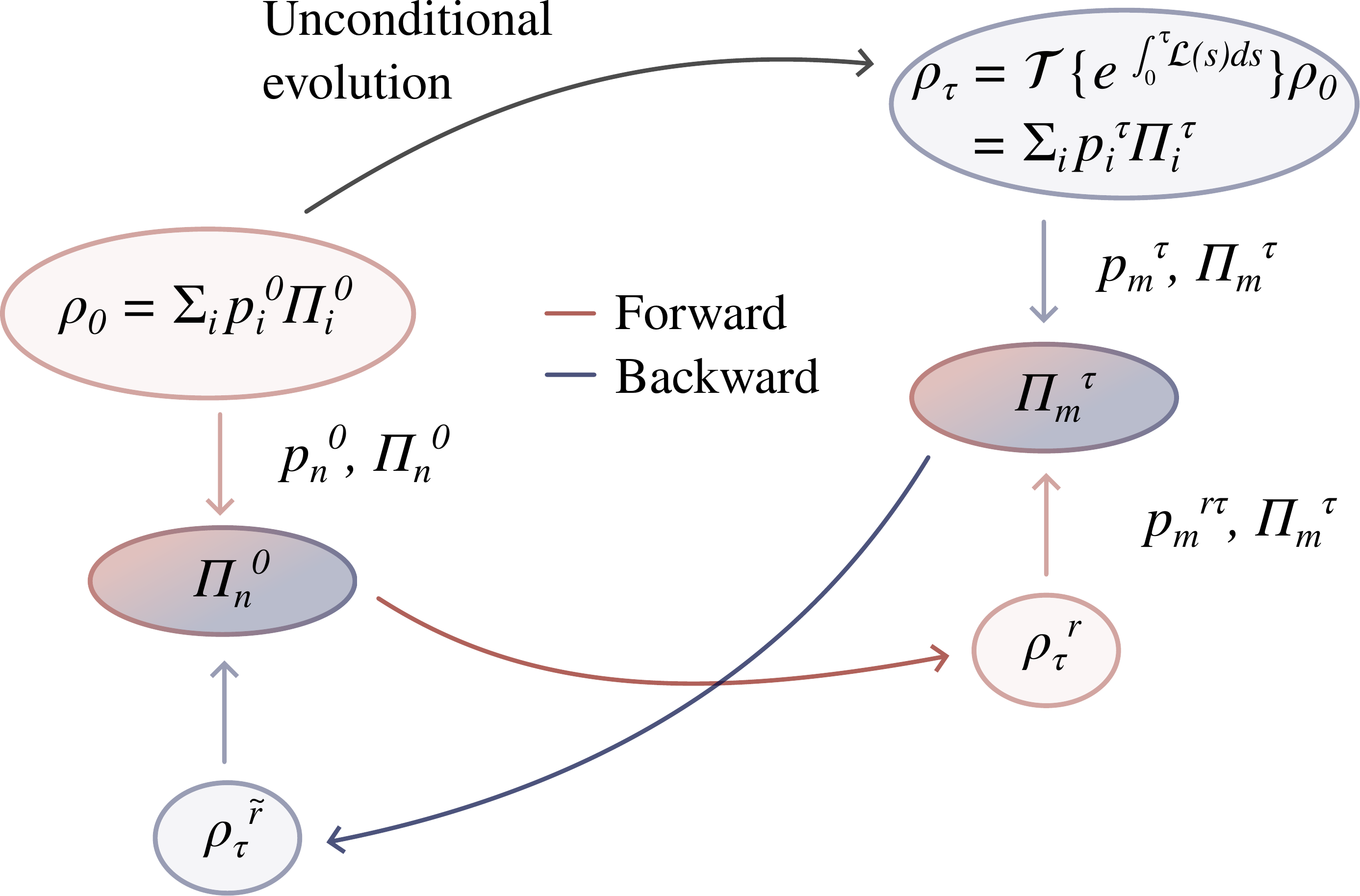}
  \caption{Schematic of the TPM scheme. In the first measurement, the initial state $\rho_0=\rho(0) $ is projected into its $n^\mathrm{th}$ eigenstate $\Pi_n^0$ with probability $p_n^0$. Subsequently, this state evolves according to the conditional evolution into $\rho_\tau^r = \rho_r(\tau)$. Finally, $\rho_r(\tau)$ is projected into the $m^\mathrm{th}$ eigenstate $\Pi_m^\tau$ of the unconditionally evolved initial state $\rho_\tau =\rho(\tau)$ with probability $p^{r\tau}_m$. The probability for the reverse trajectory to start in the eigenstate $\Pi_m^\tau$ is given by $p_m^\tau$.}
 \label{fig: TPM}
\end{center}
\end{figure}

The forward and backward trajectories can be precisely defined within the framework of the ``two-point measurement" (TPM) scheme, as schematically shown in Fig.~\ref{fig: TPM} and detailed in Ref. \cite{manzano_quantum_2022}. In the TPM scheme, the (stochastic) time-evolution of the state in the open interval $(0, \tau)$, yielding the measurement record ${r}_{\left(0,\tau\right)}$ is framed by two measurements, one at the start ($t = 0$) and one at the end of each trajectory ($t = \tau$). The full trajectory in the closed interval $\left[0,\tau \right]$, including the projective measurements, is termed the forward trajectory and we denote the corresponding measurement record by ${r}_{\left[0,\tau\right]}$. The observables measured here in the projective measurements at the beginning and end of each trajectory (see Fig.~\ref{fig: TPM}) are time-reversal invariant. In particular, we consider the case in which each of the projective measurements in the time-reversed backward process produces the same outcome as in the forward process. This means that the measurement record of the backward trajectory in between the two measurements $\Tilde{r}_{\left(0,\tau\right)}$ is exactly the time-reversed measurement record of the forward trajectory $r_{\left(0,\tau\right)}$ with $N_j$  recorded jumps, so $\Tilde{r}_{\left(0,\tau\right)} = \lbrace (\tau - t_{J_{N_j}}, k_{N_j}, - \sigma_{N_j}), (\tau - t_{J_{N_j-1}}, k_{N_j-1}, - \sigma_{N_j-1}), \dots, (\tau - t_{J_{1}}, k_{1}, - \sigma_{1}) \rbrace$
\cite{manzano_quantum_2018}.\\
We now discuss the two projective measurements. To this end, we first note that the initial state of the forward trajectory may be written in its eigenbasis $\rho(0)=\sum_i p^0_i \Pi^0_i$. If it were to evolve according to the unconditional evolution, then after time $\tau$ the system would be in state $\rho(\tau) = \mathcal{T}\left\lbrace e^{\int_0^\tau\mathcal{L}(s)\mathrm{d}s}\right\rbrace \rho(0) = \sum_i p_i^\tau \Pi^\tau_i$. The states $\rho(0)$ and $\rho(\tau)$ can be understood as the ensemble averaged initial and final state, respectively.
The initial state of each trajectory is sampled from the spectral decomposition of $\rho(0)$ via the first measurement. In particular, in the TPM the first measurmement projects $\rho(0)$ into its eigenstate $\Pi^0_n$, with probability $p^0_n$. The initial state of the trajectory is therefore $\rho_r(0) = \Pi^0_n$. This state then time-evolves up to time $\tau$ according to the conditional evolution along a trajectory defined by the measurement record. The final state of the trajectory is denoted by $\rho_r(\tau)$. Finally, at time $\tau$, there is a second projective measurement, now in the eigenbasis of the ensemble averaged final state $\rho(\tau)$ yielding $\Pi^\tau_m$ with probability $p_m^{r\tau}=\mathrm{Tr}\left[\Pi^\tau_m\rho_r(\tau)\right]$. The probability for the time-reversed trajectory to start in $\Pi^\tau_m$ is given by $p_m^\tau$, i.e. its weight in the spectral decomposition of the ensemble averaged final state $\rho(\tau)$. 

The reason we are interested in computing the projection probabilities $p_n^0$ and $p_m^\tau$ is that they appear in the expression for the total entropy production along a trajectory $r_{\left[0, \tau\right]}$ (see Eq.~\eqref{eq: S_tot}), which for the TPM scheme can be expressed as \cite{manzano_quantum_2022}
\begin{equation}
\label{eq: S_tot_2}
    S_\mathrm{tot}({r_{\left[0, \tau\right]}}) = \log\left(\frac{p^0_{n}}{p^\tau_m}\right) +\Sigma_{r_{\left(0, \tau\right)}} .
\end{equation}
Here we have split $S_\mathrm{tot}({r_{\left[0, \tau\right]}})$ into a stochastic entropy associated with the system state along a trajectory and the entropy flux transferred to the environment. The first term accounts for changes in Shannon self-information or surprisal of the extended system between the two projective measurements, and has the same form in the classical analogue \cite{crooks_entropy_1999, seifert_entropy_2005}.
The entropy flux is given by
\begin{equation}
    \Sigma_{r_{\left(0, \tau\right)}}= -\sum_{\alpha=1}^{N_\mathrm{R}} \sum_{k=1}^{l_\alpha} \sum_{\sigma \in \lbrace+,-\rbrace}  \int_0^\tau \frac{\Delta E^\sigma_{k,\alpha} - \mu_\alpha \Delta N^\sigma_{k,\alpha}}{T_\alpha}  \mathrm{d}N_k(t),
\end{equation}
where $\Delta E_{k, \alpha}$ is assumed to be the energy of the lead mode $k$ in lead $\alpha$, as discussed also Sec.~\ref{sec: meas_work}. A discussion of entropy production and fluctuation theorems accounting for imperfect detection may be found in Ref. \cite{ferri-cortes_entropy_2023}.\\

\subsection{Efficient computation of $p^0_{n}$, $C^r(0)$ and $p^\tau_m$ }
\label{sec:computing_probs}

In the following, we discuss how the projection probabilities $p^0_{n}$ and $p^\tau_m$ as well as the covariance matrix computed with respect to the state after the first projective measurement $C^r(0)$ may be obtained efficiently.

\subsubsection{Probability for the forward trajectory to start in $\ket{s_n}$}
\label{sec: F_P1}
The initial state of the extended system $\rho(0)$ is Gaussian, therefore it may be expressed as
\begin{equation} \begin{split}
    \rho(0) &= \frac{e^{-\mathbf{c}^ \dagger  M(0) \mathbf{c}}}{Z(0)}, \hspace{1cm} \mathbf{c} = (c_1, c_2, \dots, c_{N_\mathrm{S} + L})\\
    &= \prod_{i=1}^{N_\mathrm{S} + L}\frac{e^{-\mu^0_i d_i^ \dagger  d_i}}{Z_i(0)},
 \end{split}\end{equation}
where $\mu^0_i$ are the eigenvalues of $M(0)$, and $Z_i(0) = 1+ e^{-\mu^0_i}$. The $2^{N_\mathrm{S} + L}$ eigenvalues of $\rho(0)$ are of the form 
\begin{equation}
    p_n^0 = \prod_{i=1}^{N_\mathrm{S} + L} \frac{e^{-\mu^0_i s_n^i}}{Z_i(0)}, 
\end{equation}
where $s_n = s_n^1s_n^2\dots s_n^{N_\mathrm{S} + L}$ is a binary string, with $s_n^i \in \lbrace  0,1 \rbrace$. Note that $\mu^0_i = \left(U^ \dagger  M(0) U\right)_{ii}$, where $U$ diagonalises the covariance matrix of the initial state $C(0)$. 
The $2^{N_\mathrm{S} + L}$ eigenvectors of $\rho(0)$ are 
\begin{equation}
    \ket{s_n} = \prod_{i=1}^{N_\mathrm{S} + L} (d_i^ \dagger )^{s_n^i} \ket{0}.
\end{equation}
Note here that fermionic Fock states are Gaussian as they are connected to the vacuum by a Gaussian unitary
\begin{equation}
    \ket{s_n} = U_{s} \ket{0}, \hspace{1cm} U_{s} = \prod_{i=1}^{N_\mathrm{S} + L} \left(d_i^ \dagger  d_i\right)^{s_n^i},
\end{equation}
and
\begin{equation}
    d_j^\prime = U_s d_j U_s^ \dagger  = \begin{cases}
        \pm d_j \hspace{0.5cm} (s_{m}^j = 0)\\
        \pm d_j^ \dagger  \hspace{0.5cm} (s_m^j = 1),\\
    \end{cases}
\end{equation}
where $\pm$ stands for a phase factors and is not relevant for whether state is Gaussian or not.
We then find
\begin{equation}
    \lbrace  d_j^\prime, {d_k^\prime}^ \dagger \rbrace = \delta_{jk},
\end{equation}
and therefore $U_s$ is Gaussian.
The probability of projecting $\rho(0)$ onto its eigenstate $\ket{s_n}$ during the first projective measurement is determined by the corresponding eigenvalue  $p^0_n$.
\subsubsection{Covariance matrix after the first projective measurement}
Given the covariance matrix $C(0)$ computed with respect to ensemble average state $\rho(0)$, one constructs the diagonalising unitary $U_0$, such that
\begin{equation}
    C(0) = U_0 \Lambda U_0^ \dagger ,
\end{equation}
where $\Lambda = \mathrm{diag}(\mathrm{eigenvals(C(0))})$. 
The covariance matrix $C^r(0)$, computed with respect to the eigenstate $\ket{s_n}$ of $\rho(0)$, is given by
\begin{equation}
    C^r(0) = U_0 n_{s_n} U_0^ \dagger ,
\end{equation}
where $(n_{s_n})_{ij} = \bra{s_n}d_j^\dagger d_i\ket{s_n}$, as detailed in Appendix~\ref{sec: connection_C_n_s}. Note that the set of matrices $\lbrace n_{s_i}\rbrace$ is simply given by the set of all $2^{N_\mathrm{S} + L}$ diagonal matrices with bit-strings of length $N_\mathrm{S} + L$ (with letters either 0 or 1) on their diagonal. For instance, for $N_\mathrm{S} + L=2$: $\lbrace n_{s_i} \rbrace = \lbrace  \mathrm{diag}\left((0,0)\right), \mathrm{diag}\left((0,1)\right), \mathrm{diag}\left((1,0)\right), \mathrm{diag}\left((1,1)\right)\rbrace$.

\subsubsection{Probability to project into $m^\mathrm{th}$- eigenstate of $\rho(\tau)$}

The ensemble average final state of the extended system $\rho(\tau)$ is Gaussian, therefore it may be expressed as
\begin{equation} \begin{split}
    \rho(\tau) &= \frac{e^{-\mathbf{c}^ \dagger  M(\tau) \mathbf{c}}}{Z(\tau)}, \hspace{1cm} \mathbf{c} = (c_1, c_2, \dots, c_{N_\mathrm{S} + L})\\
    &= \prod_{i=1}^{N_\mathrm{S} + L} \frac{e^{-\mu^\tau_i d_i^ \dagger  d_i}}{Z_i(\tau)},
 \end{split}\end{equation}
where $\mu^\tau_i$ are the eigenvalues of $M(\tau)$, and $Z_i(\tau) = 1+ e^{-\mu^\tau_i}$. The eigenvalues of $\rho(\tau)$ are of the form 
\begin{equation}
\label{eq: p_m_tau}
    p_m^\tau = \prod_{i=1}^{N_\mathrm{S} + L} \frac{e^{-\mu^\tau_i s_m^i}}{Z_i(\tau)}, 
\end{equation}
and the eigenvectors of $\rho(\tau)$ are given by
\begin{equation}
    \ket{s_m} = \prod_{i=1}^{N_\mathrm{S} + L} (d_i^ \dagger )^{s_m^i} \ket{0}.
\end{equation}
In the following we are interested in computing the probability of the final state of the trajectory $\rho_r(\tau)$ to be projected into the eigenstate $\ket{s^o_m}$ (here in the original/operational basis) of the ensemble average density matrix of the final state $\rho(\tau)$, with $\ket{s^o_m}\bra{s^o_m}= \Pi_m^\tau$,
\begin{equation}
    p^{r\tau}_m = \mathrm{Tr}\left[\rho_r(\tau) \ket{s^o_m}\bra{s^o_m}\right].
\end{equation}
In particular, we are interested in computing it given only the covariance matrix $C(\tau)$ of the ensemble average of the final state $\rho(\tau)$ and the covariance matrix $C^r(\tau)$ of the final state of the trajectory $\rho_r(\tau)$.

Let us first write $\rho_r(\tau)$ in the eigenmode basis of $\rho(\tau)$
\begin{equation}
    \rho^\prime_r(\tau) = \frac{e^{-\mathbf{d}^ \dagger  Q \mathbf{d}}}{Y},
\end{equation}
so that
\begin{equation} \begin{split}
    p^{r\tau}_m 
    &= \mathrm{Tr}\left[\rho^\prime_r(\tau) \ket{s_m}\bra{s_m}\right].
 \end{split}\end{equation}
Using the functional determinant formula~\cite{abanin_fermi-edge_2005, klich_full_2002}, which maps a many-body expectation value onto a determinant in single-particle space, we find
\begin{equation} \begin{split}
    p^{r\tau}_m 
    &= \frac{1}{Y} \mathrm{det}\left[1- n_{s_m} + n_{s_m} e^{-Q} \right],
 \end{split}\end{equation}
where $\ket{s_m} = U\ket{s^o_m}$ and
\begin{equation}
    n_{s_m} = \mathrm{diag}\left((s_m^1, \dots, s_m^{N_\mathrm{S} + L})\right).
\end{equation}
The above expression can be further simplified, when remembering that $Q$ and $Y$ may be expressed in terms of the covariance matrix $C^{r\prime}(\tau)$ expressed in the eigenbasis of $C(\tau)$, to circumvent numerical complications arising from having eigenvalues of either 0 or 1 in their spectrum,
\begin{equation} \begin{split}
    Q &= \log\left[\frac{1-C^{r\prime}(\tau)}{C^{r\prime}(\tau)}\right],\\
    Y &= \det\left[\frac{1}{1-C^{r\prime}(\tau)}\right],
 \end{split}\end{equation}
so that 
\begin{equation} \begin{split}
    p^{r\tau}_m 
    & = \det\left[(1-n_{s_m})(1-C^{r\prime}(\tau)) + n_{s_m} C^{r\prime}(\tau)\right].
 \end{split}\end{equation}
 
\subsubsection{Probability for the backward trajectory to start in $\ket{s_m}$}

Now, what remains to be understood is the probability for the reverse trajectory to start in the eigenstate $\ket{s_m}$ of the ensemble-averaged final state $\rho(\tau)$. This probability is simply given by the eigenvalue $p_m^\tau$ of $\rho(\tau)$ associated with the eigenstate $\ket{s_m}$, defined in Eq.~\eqref{eq: p_m_tau}, and can thus be computed following the same method as described in Sec.~\ref{sec: F_P1}.
\subsection{Uncertainty and martingale entropy production}

As is well known, there exist multiple decompositions of the total entropy production into physically meaningful contributions~\cite{Esposito2010,manzano_quantum_2019}. Instead of splitting the total entropy production into terms arising from the change of the system state and from dissipation into the environment, one may split $S_{\rm tot}$ into two contributions that explicitly quantify the effect of measurement backaction on entropy production within the TPM scheme \cite{manzano_quantum_2022, manzano_quantum_2019}. 

Explicitly, we can write
\begin{equation}
    S_\mathrm{tot}({r_{\left[0, \tau\right]}}) = S_\mathrm{unc}({r_{\left[0, \tau\right]}}) + S_\mathrm{mart}({r_{\left[0, \tau\right]}}).
\end{equation}
The uncertainty entropy production is defined by
\begin{equation}
\label{S_unc}
    S_\mathrm{unc}({r_{\left[0, \tau\right]}}) = -\log(p^\tau_m) - S_\rho(\tau),
\end{equation}
where $ S_\rho(\tau) = -\log\left(\mathrm{Tr}\left[\rho_r(\tau) \rho(\tau)\right]\right)$ and  $\rho_r(\tau)$ is the state conditioned on the measurement record prior to the second projective measurement, while $\rho(\tau)$ is the unconditional final state. Eq.~\eqref{S_unc} measures how much information we gain knowing the outcome of the second projective measurement $m(\tau)$ --- and thus $\rho_r(\tau)$ --- relative to knowing only the unconditional state $\rho(\tau)$. The martingale entropy production, named for its characteristic of being an exponential martingale along quantum trajectories \cite{manzano_quantum_2019}, is given by
\begin{equation}
    S_\mathrm{mart}({r_{\left[0, \tau\right]}}) = \log(p^0_n) + S_\rho(\tau) + \Sigma_{r_{\left(0, \tau\right)}}.
\end{equation}
Both contributions to the total entropy production fulfill an integral fluctuation theorem, so that $\left \langle e^{-S_\mathrm{unc}({r_{\left[0, \tau\right]}})}\right\rangle_r = 1$ and $\left \langle e^{-S_\mathrm{mart}({r_{\left[0, \tau\right]}})}\right\rangle_r = 1$, and therefore are non-negative on average, i.e. $\left \langle S_\mathrm{unc}({r_{\left[0, \tau\right]}})\right\rangle_r \geq 0$ and $\left \langle S_\mathrm{mart}({r_{\left[0, \tau\right]}})\right\rangle_r \geq 0$ \cite{manzano_quantum_2022}. 

Interestingly, this splitting allows one to describe classical and quantum contributions at the level of single trajectories separately. $S_\mathrm{unc}(\tau)$ identifies the entropy production related to the second projective measurement on the system --- thus is due to the intrinsic quantum uncertainty --- and vanishes in classical settings, where the system is always in one of its eigenstates. Importantly, the uncertainty entropy production is non-extensive in time, whereas the martingale entropy production may be extensive in time, for example in non-equilibrium steady states, because of its dependence on the entropy flow due to a net heat exchange with the environment.

In Sec.~\ref{sec:computing_probs} we have discussed how to compute $p_n^0$, $p_m^\tau$, as well as $\Sigma_{r_{\left(0, \tau \right)}}$ in terms of covariance matrices, given the state is Gaussian fermionic at all times along the trajectory. Also $S_\rho(\tau)$ can be computed relying soley on covariance matrices, since for two Gaussian fermionic states $\rho_1 = e^{-\mathbf{c}^\dagger M_1 \mathbf{c}}/Z_1$ and $\rho_2= e^{-\mathbf{c}^\dagger M_2 \mathbf{c}}/Z_2$, using using the Baker-Campbell-Hausdorff formula, so that
\begin{align}
    e^{-\mathbf{c}^\dagger M_1 \mathbf{c}}e^{-\mathbf{c}^\dagger M_2 \mathbf{c}}  = & -\mathbf{c}^\dagger M_1 \mathbf{c} - \mathbf{c}^\dagger M_2 \mathbf{c}  \notag \\ &   -\mathbf{c}^\dagger \left[M_1, M_2\right] \mathbf{c} + \dots,
\end{align}
one can then show that 
\begin{equation}
\begin{split}
    e^{-\mathbf{c}^\dagger M_1 \mathbf{c}}e^{-\mathbf{c}^\dagger M_2 \mathbf{c}} & = e^{-\mathbf{c}^\dagger \chi \mathbf{c}}\\
\end{split}
\end{equation}
with
\begin{equation}
\begin{split}
    \chi & = \log\left(e^{M_1} e^{M_2}\right).
\end{split}
\end{equation}
Using the relation
\begin{equation}
    \mathrm{Tr}\left[e^{-\mathbf{c}^\dagger \chi \mathbf{c}}\right] = \mathrm{det}\left[1+e^{-\chi}\right],
\end{equation}
we find
\begin{equation}
\begin{split}
     \mathrm{Tr}\left[\rho_1\rho_2\right]&= \frac{\mathrm{det}\left(1+e^{-M_1}e^{-M_2}\right)}{Z_1Z_2}.\\
\end{split}
\end{equation}

\subsection{Verification of the integral fluctuation theorems for entropy production}
\label{sec: val_FT}

\begin{figure}[t]
\begin{center}
\includegraphics[width=0.3\linewidth]{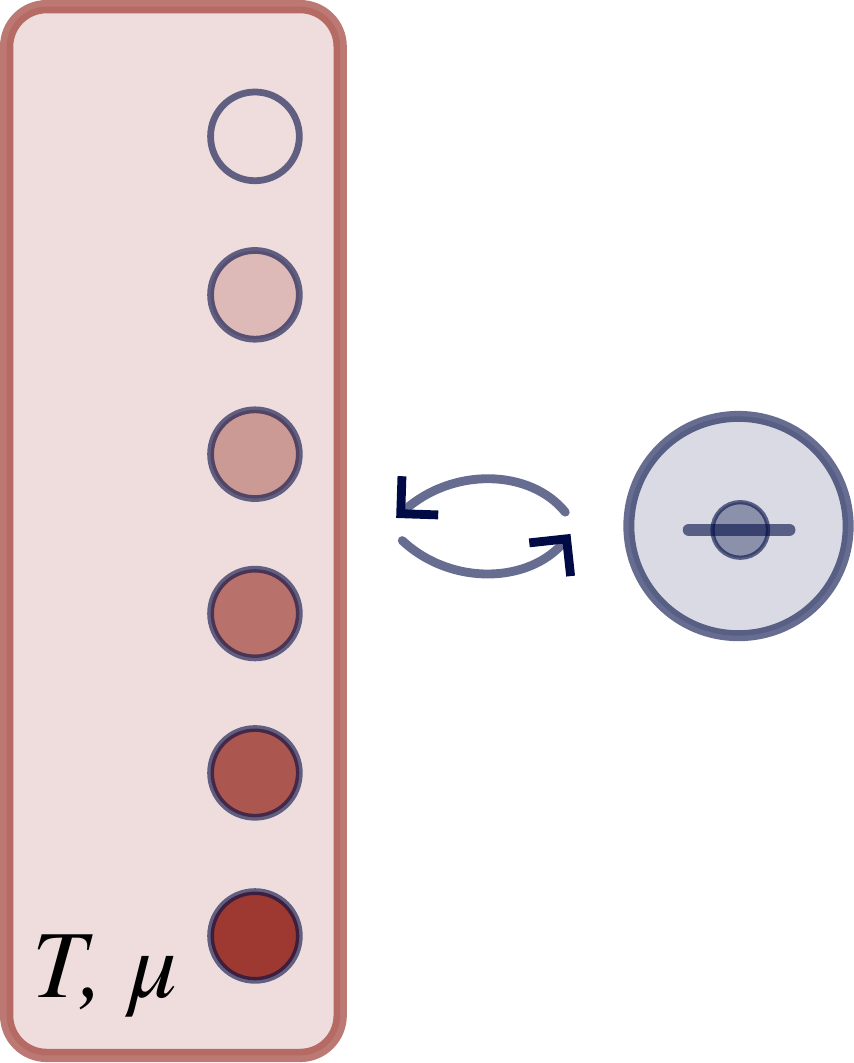}
  \caption{Schematic of a single dot coupled to a fermionic reservoir in the mesoscopic-leads formalism.}
 \label{fig: single_dot_schematic}
\end{center}
\end{figure}

We now consider a set-up in which
a single quantum dot with Hamiltonian 
\begin{equation}
    H_\mathrm{S} = \epsilon f^\dagger f
\end{equation}
is coupled to a fermionic reservoir, represented in the mesoscopic-leads formalism, as shown in Fig. \ref{fig: single_dot_schematic}, in the steady-state.
For simplicity, we consider the case in which the reservoir is described by a flat spectral density, given by
\begin{equation}
    J(\omega) = \begin{cases}
        \Gamma & \omega \in \left[-\omega_\mathrm{max}, \omega_\mathrm{max}\right],\\
        0 & \text{otherwise,}
    \end{cases}
\end{equation}
where $\Gamma$ denotes the coupling strength between the quantum dot and the reservoir and $\omega_\mathrm{max}$ is a hard cut-off.
We choose the set of lead mode energies $\lbrace\epsilon_k\rbrace$, the couplings between the quantum dot and the lead modes $\lbrace\kappa_k\rbrace$ and the damping rates of the residual reservoirs $\lbrace\gamma_k\rbrace$ as described in Sec.~\ref{sec: mesoleads_formulation}. Here, we choose the energies $\lbrace\epsilon_k\rbrace$ via linear discretization.  For details on the construction of the Liouvillian governing the dynamics of the the extended system given these parameters, we refer to Sec.~\ref{sec: mesoleads_formulation}. Trajectories are recorded using the TPM scheme, with projective measurements taken in the eigenbasis of the steady-state of the extended system, both at the start and end of each trajectory, as described in Sec.~\ref{sec: EP_TPM}.

\begin{figure}[t]
\begin{center}
\includegraphics[width=\linewidth]{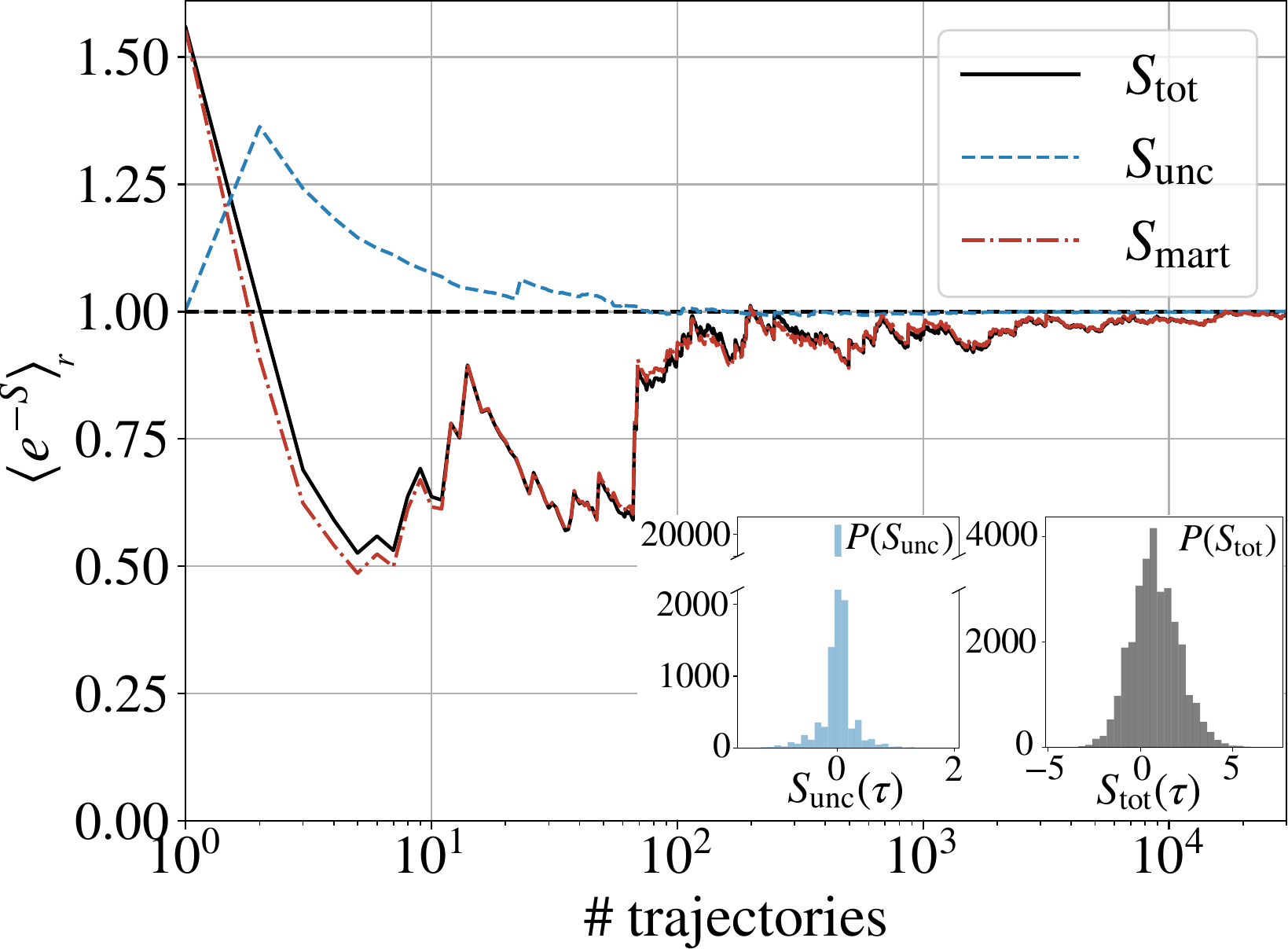}
  \caption{Verification of the integral fluctuation theorem in the TPM scheme for the total entropy production (black), the uncertainty entropy production (blue) and the martingale entropy production (red) in the steady-state. The insets show the distribution of $S_\mathrm{tot}$ (left) and $S_\mathrm{unc}$ (right). The estimated probability densities $P(S_\mathrm{tot})$ and $P(S_\mathrm{unc})$, respectively, are obtained by dividing the number of counts over the total number of trajectories. Parameters:
  $L$ = 10, up to 30000 trajectories, $\omega_\mathrm{max} = 1$, $\epsilon$ = $\omega_\mathrm{max}/4$, $\Gamma = \omega_\mathrm{max}/8$, $T =\omega_\mathrm{max}$, $\mu = \omega_\mathrm{max}/16$, $\Gamma\tau = 50$.}
 \label{fig: FT_test}
\end{center}
\end{figure}

\begin{figure*}[t]
\begin{center}
\includegraphics[width=\linewidth]{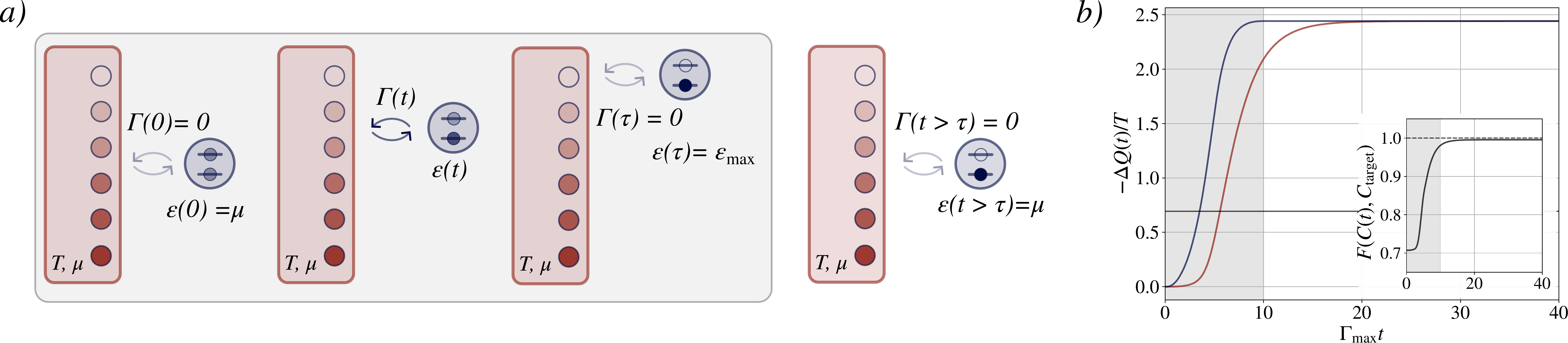}
  \caption{
  a) Driving protocol of both site energy $\epsilon(t)$ and coupling strength $\Gamma(t)$: the equilibration period (outside the grey shaded region) allows equilibration of the lead modes with respect to their residual reservoirs. b) Time evolution of the dissipated heat from the system (blue) and the extended system (red) during the execution of the driving protocol and equilibration time. As we allow for equilibration of the lead modes to their residual reservoirs after uncoupling from the system site, they agree. The grey line indicates Landauer's bound.  The inset shows the time evolution of the fidelity to the target state. }
 \label{fig: erasure_example}
\end{center}
\end{figure*}

We find that the integral fluctuation theorems for $S_\mathrm{tot}({r_{\left[0, \tau\right]}})$, $S_\mathrm{unc}({r_{\left[0, \tau\right]}})$ and $S_\mathrm{mart}({r_{\left[0, \tau\right]}})$ are verified individually for their respective distributions after the second measurement, since the functionals $\left\langle e^{-S_\mathrm{tot}({r_{\left[0, \tau\right]}})}\right\rangle_r$, $\left\langle e^{-S_\mathrm{unc}({r_{\left[0, \tau\right]}})}\right\rangle_r$ and $\left\langle e^{-S_\mathrm{mart}({r_{\left[0, \tau\right]}})}\right\rangle_r$ converge to 1 as the number of trajectories employed in the simulations increases, see Fig. \ref{fig: FT_test}. Notably, convergence to the uncertainty entropy production fluctuation theorem is reached the quickest, in agreement with results presented in \cite{manzano_quantum_2022}.
In all cases, approximate convergence occurs after roughly 100 trajectories, a typical scale for the convergence of the unconditional density operator in conventional quantum trajectory simulations. However, complete convergence of $S_{\rm tot}$ and $S_{\rm mart}$ is not achieved until two orders of magnitude more trajectories, as we are examining a rather detailed statistical property of the reservoir, requiring sufficient sampling of tails of the distributions. 

By demonstrating the validity of these fluctuation theorems for entropy production, we have established the thermodynamic consistency of the mesoscopic-leads approach at the stochastic level. This represents one of the key results of our work. Our framework naturally incorporates measurement conditioning at the trajectory level, allowing us to also assess how entropy production arises from uncertainty and martingale contributions. 

The inset of Fig.~\ref{fig: FT_test} shows the estimated distributions of the total entropy production and uncertainty entropy production, $P(S_\mathrm{tot})$ and $P(S_\mathrm{unc})$, respectively.
The total entropy production appears to be close to a shifted Gaussian with positive mean, verifying the second law inequality. By contrast, the uncertainty entropy production instead shows a large peak significantly closer to zero, with secondary peaks on both sides of this central peak. We emphasise that the ability to directly sample and visualise these detailed features of the distribution is a key advantage of our method.

We now briefly address why entropy production along trajectories remains nonzero and its average does not vanish, even in this example of an equilibrium process without external driving. This arises from the definition of total entropy production (Eq.~\eqref{eq: S_tot_2}) under a local GKSL master equation, where bath-induced transitions do not mediate between system eigenstates. As a result, a given relative surprisal in measurement outcomes ($\log p_n^0/p_m^\tau$) cannot be uniquely linked to a specific heat change, unlike in a global GKSL framework. Consequently, even in an equilibrium state, there exists a distribution of total entropy flux. The nonzero average stems from the exclusion of measurement energy (Eq.~\eqref{eq: I_EM}) in Eq.~\eqref{eq: S_tot_2}. However, including it would violate the integral fluctuation theorem in Eq.~\eqref{eq: integral_FT}\cite{manzano_quantum_2022}. The fact that the average measurement energy (Eq.~\eqref{average_measurement_energy}) is nonzero is a well-known issue in the mesoscopic leads framework and has been discussed before in Refs.~\cite{brenes_tensor-network_2020,  Lacerda2023, lacerda_entropy_2023}. However, this small discrepancy vanishes as the number of lead modes increases. We have included a more detailed discussion of this feature of the method in Appendix~\ref{sec: FT_appendix}.

\section{Heat dissipation fluctuations in finite-time information erasure}
\label{sec: heat_ft_le}
\begin{figure*}[t]
\begin{center}
\includegraphics[width=\linewidth]{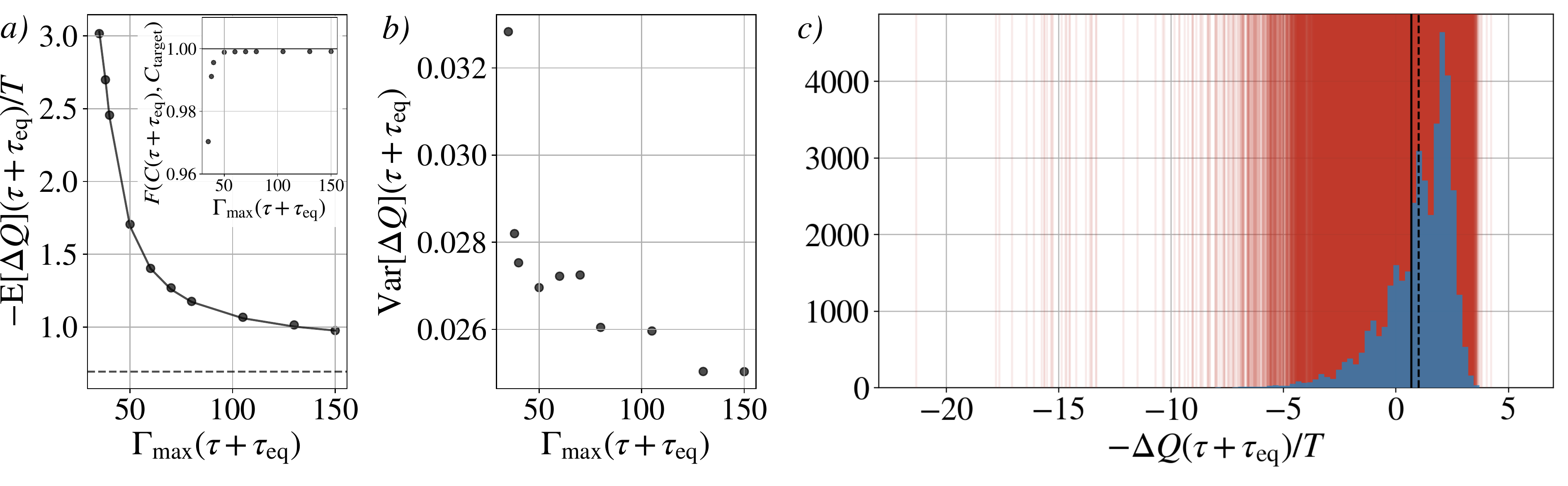}
  \caption{The mean a) and variance b) of the distribution of dissipated heat during the information erasure protocol both decrease as a function of the protocol duration. The dotted line in a) indicates the Landauer bound. The black dots represent data derived from the distribution of dissipated heat obtained through trajectory sampling. In contrast, the line plot illustrates the results obtained when considering the unconditional evolution of the system subject to the erasure protocol. The inset shows the quantum fidelity to the target state as a function of the driving duration. c) Trajectory-sampling of the distribution of dissipated heat (blue) reveals non-Gaussian statistics, even at slow-driving conditions with $\Gamma_\mathrm{max} \tau =100$. The red vertical lines show the dissipated heat along the individual trajectories. The distribution's mean is represented by the black dashed line, while the Landauer bound is indicated by the solid line. 
  Parameters: 40000 trajectories, $L = 20$, $\omega_\mathrm{max} = 1$, $T = \omega_\mathrm{max}/10 $, $\mu = - 0.8 \omega_\mathrm{max}$, $\epsilon_\mathrm{max} = 0.8\omega_\mathrm{max}$, $\Gamma_\mathrm{max} = \epsilon_\mathrm{max}/\pi$, $\Gamma_\mathrm{max}\tau_\mathrm{eq} = 30$.
 }
 \label{fig: dQ_varQ_fid}
\end{center}
\end{figure*}

Having verified the thermodynamic consistency of our approach, we now apply it to study heat fluctuations in the paradigmatic example of finite-time information erasure. Landauer’s principle asserts a fundamental limit to the thermodynamic cost of erasing information:
\begin{equation}
\label{landauer_bound}
    -\Delta Q \geq T\log(2),
\end{equation}
where $-\Delta Q$ is the heat dissipated into the bath during the erasure process. The limit can be saturated only by a reversible and isothermal process, which requires infinite time. Recently, several studies have unveiled corrections to Landauer's principle that appear in finite-time protocols~\cite{miller_quantum_2020,Zhen2021,Vu2022,Dago2022,rolandi_finite-time_2022}, but these studies have mostly been limited to the regime of weak system-bath coupling or slow driving. 

Here, we exploit our method to study how varying the driving speed of the process impacts the dissipated heat fluctuations during information erasure with strong system-bath coupling. We consider a bit of information encoded in the occupation of a single fermionic mode (bit mode). The bit is erased by manipulating its time-dependent Hamiltonian,
$H_\mathrm{S}(t) = \epsilon(t)f^\dagger f$, while in contact with a heat bath at temperature $T$ and chemical potential $\mu$. As above, we model this heat bath by a mesoscopic lead, as prescribed in Sec.~\ref{sec: mesoleads_formulation}. 

Initially the bit mode and the lead modes are uncorrelated. The bit mode has occupation $\frac{1}{2}$ and the lead modes are populated according to the Fermi-Dirac distribution $f(\epsilon, T, \mu)$.
The aim is to reach the target state in which the bit mode population is reduced to 0 and the lead modes are again uncorrelated and populated according to $f(\epsilon, T, \mu)$. If the protocol is successful, one bit of information has been erased. 
To this end an external drive is  applied, so that the energy of the mode changes according to
\begin{equation}
    \epsilon(t) = \mu + (\epsilon_\tau - \mu)\left(\frac{t}{\tau} - \frac{\sin(2\pi t/\tau)}{2\pi}\right),
\end{equation}
where $\epsilon_\tau$ denotes the energy splitting of the bit mode at the final time.
The overall coupling $\Gamma$ also changes over time as
\begin{equation}
    \Gamma(t) = \frac{\epsilon_\tau}{\pi} \sin^2\left(\pi \frac{t}{\tau}\right).
\end{equation}
Here, we assume a flat spectral density
\begin{equation}
    \mathcal{J}(\omega_k) = \begin{cases} 
\Gamma(t) & \omega_k \in \left[-\omega_\mathrm{max}, \omega_\mathrm{max}\right] \\ 
0 & \text{otherwise},
\end{cases}
\end{equation}
where $\omega_\mathrm{max}$ is a hard cut-off.
We linearly discretise the reservoir into $L$ energy modes so that $\Delta\omega = \omega_{k+1} - \omega_k = 2\omega_\mathrm{max}/L$ and choose the damping rate $\gamma_k = \Delta \omega$.
The time-dependent coupling strength between the bit mode and the $k$-th lead modes is given by $\kappa_k(t) = \sqrt{\frac{\Gamma(t) \Delta\omega}{2\pi}}$.

After the driving protocol is executed, the leads are left to equilibrate for an additional time $\tau_\mathrm{eq}$. This ensures that the erasure protocol describes a cycle, resetting the lead modes to their initial state, where the lead modes are uncorrelated and populated according to the Fermi-Dirac distribution $f(\epsilon, T, \mu)$.  Further, it ensures that the total average heat dissipated out of the extended system and into the residual reservoirs, defined as 
\begin{equation}
    \Delta Q^r(t, 0) = \Delta E^r(t, 0) - \mu\Delta N^r(t, 0),
\end{equation}
matches the total average heat dissipated from the charge bit into the lead modes, as shown in Fig.~\ref{fig: erasure_example} b), given by
\begin{equation}
    \Delta Q^r_S(t, 0) = \int_0^t \mathrm{d}t^\prime \left[ J_E(t^\prime) - \mu J_N(t^\prime) \right],
\end{equation}
where the so-called internal average particle and energy currents in the mesoscopic-leads formalism, exchanged between the system and the lead modes in lead $\alpha$, are generally defined as \cite{lacerda_quantum_2022}
\begin{align}
    J_\alpha^N(t) &= i\left\langle \left[ N_{\mathrm{L}_\alpha}, H_{\mathrm{SL}_\alpha}\right]\right \rangle,\\
    J_\alpha^E(t) &= i\left\langle \left[H_{\mathrm{L}_\alpha}, H_{\mathrm{SL}_\alpha}\right]\right \rangle + \mathrm{Tr}\left[H_{\mathrm{SL}_\alpha} \mathcal{L}_\alpha \left[\rho\right]\right],
\end{align}
respectively, where $N_{\mathrm{L}_\alpha} = \sum_{k=1}^{l_\alpha} a_{\alpha, k}^\dagger a_{\alpha, k}$, $J^E(t) = \sum_{\alpha = 1}^{N_\mathrm{R}} J_\alpha^E(t)$ and $J^N(t) = \sum_{\alpha = 1}^{N_\mathrm{R}} J_\alpha^N(t)$.

We find that the mean dissipated heat approaches Landauer's bound when the driving speed is reduced, as expected
(see Fig.~\ref{fig: dQ_varQ_fid} a)). 
Further, the quantum fidelity to the target state, which for fermionic Gaussian states is given by \cite{lacerda_entropy_2023, banchi_quantum_2014}
\begin{equation}
    F(\rho_1 || \rho_2) = \mathrm{Tr}\left[\sqrt{\rho_1 \rho_2}\right] = \frac{\mathrm{det}\left(\mathbb{1}+e^{-M_1/2} e^{-M_2/2}\right)}{\sqrt{Z_1 Z_2}},
\end{equation}
approaches 1, as shown in the inset of Fig.~\ref{fig: dQ_varQ_fid} a).
While it is known how to compute fluctuations in the (internal) particle current in the context of mesoscopic leads using full counting statistics \cite{brenes_particle_2023}, this is not the case (yet) for fluctuations in the (external) energy and heat currents. In the approach we take here, we reconstruct the distribution of dissipated heat by sampling from it, and we find (the trend) that also the variance of the dissipated heat decreases as $\tau$ increases, as shown in Fig.~\ref{fig: dQ_varQ_fid} b). 

Fig.~\ref{fig: dQ_varQ_fid} c) shows a histogram of the heat distribution obtained by trajectory sampling, which reveals distinctly non-Gaussian statistics even under slow-driving conditions. Previous work has shown that non-Gaussian heat statistics appear during slow information erasure in the weak-coupling regime, whenever quantum coherence is generated along the protocol~\cite{miller_quantum_2020,Vu2022}. Here, coherence is expected due to the strong system-bath interaction; however, in contrast to the findings in Ref.~\cite{miller_quantum_2020}, we observe a negative skewness in the heat distribution. In addition, we observe rare events with extremely large heat flow into the system, despite the majority of trajectories yielding heat flow into the bath as demanded by Ineq.~\eqref{landauer_bound}. It is important to note that the model of the bath employed here differs fundamentally with that of Ref.~\cite{miller_quantum_2020} and thus agreement with those results is not to be expected. In particular, the dissipators and the Hamiltonian do not commute even though together they bring the system to equilibrium for large enough leads~\cite{lacerda_entropy_2023}. Fig.~\ref{fig: dQ_varQ_fid} c) thus underlines the novel and nontrivial features of heat fluctuations that can emerge at strong system-bath coupling. 

We also note that the heat distribution in Fig.~\ref{fig: dQ_varQ_fid} c) exhibits fine structure that reflects the discrete nature the lead. These features can be eliminated either by increasing the number of lead modes or via a coarser binning procedure. The number of lead modes thus sets the basic limit of energy resolution within our method.

\section{Discussion}
 \label{sec: conclusion}

The development of quantum stochastic thermodynamics has often closely followed the classical theory, where the notion of the trajectory is paramount. To make sense of trajectories in the quantum regime, one typically defines a trajectory as a sequence of classical measurement outcomes, which are (or, in principle, could be) recorded by a detector. A continuous-time description of quantum stochastic thermodynamics therefore must invoke the theory of continuous weak measurements~\cite{manzano_quantum_2022}, but such measurements lead to Markovian open quantum system dynamics by their very nature. This issue has limited previous work in this direction to the regime of weak system-environment coupling. 
 
In this work, we have extended the continuous-time trajectory description of quantum stochastic thermodynamics to the strong coupling regime by exploiting a Markovian embedding. In particular, we have presented a comprehensive analysis of the conditional dynamics for the mesoscopic-leads approach, applied to non-interacting fermionic systems. We explicitly allow for imperfect detection and partial monitoring, enabling the exploration of various measurement configurations. Our analysis, based on the framework of quantum stochastic thermodynamics, yields thermodynamically consistent analytic expressions for particle and energy currents exchanged between the extended system and the residual reservoirs. We emphasize that these external currents correspond to those observable in experiments. They remain valid even in the strong coupling regime, where the differentiation between internal and external currents becomes pronounced. We emphasise that testing these predictions would not require an experimenter to identify or distinguish the lead from the residual reservoirs: these are simply theoretical constructs. Experimentally, one need only measure the integrated current with an ammeter and compare with the integrated current calculated using our method.

To validate the thermodynamic consistency of our formalism, we have shown that integral fluctuation theorems for total entropy production, uncertainty entropy production, and martingale entropy production hold. We note that the Markovian embedding enables us to circumvent known thermodynamic inconsistencies associated with conventional GKLS MEs~\cite{carmichael_master_1973,wichterich_modeling_2007,rivas_markovian_2010,levy_local_2014,barra_thermodynamic_2015,trushechkin_perturbative_2016,gonzalez_testing_2017,hofer_markovian_2017,naseem_thermodynamic_2018,chiara_reconciliation_2018,mitchison_non-additive_2018,tupkary_fundamental_2022}. Finally, we have applied the formalism to the example of finite-time erasure of a bit of information stored in a charge qubit. By extracting fluctuations of dissipated heat while varying the driving speed, we have emphasized the method's applicability beyond the slow-driving regime. Further, we have demonstrated the method's capability to resolve the non-Gaussian statistics of the dissipated heat, even in the slow-driving regime.

The approach presented here relies solely on the covariance matrix of the extended system, enabling highly efficient numerical computations for non-interacting fermionic systems. However, our general expressions for stochastic currents hold for arbitrary systems, and could be applied to interacting problems when combined with a tensor-network representation of the state of the extended system~\cite{brenes_tensor-network_2020}. While sampling the eigenvalues of the density matrix in such a tensor-network representation would be challenging, this step was needed only to evaluate the stochastic entropy production and verify the fluctuation theorems. Conversely, trajectory sampling of charge and energy currents requires only the application of local operations within the mesoscopic-leads formulation and is thus straightforward for tensor-network theory, in principle. Moreover, while we considered reservoirs with a flat spectral density here for simplicity, it is important to highlight that our method is versatile and can be extended easily to more complex reservoirs with a non-trivial spectral density. Our work thus paves the way to implementing the full toolbox of quantum measurement-based control within a consistent thermodynamic framework for mesoscopic systems.

\vspace*{1.5cm}
 \begin{acknowledgements}
 We thank Artur Lacerda for fruitful discussions and feedback. LPB and JG acknowledge SFI for support through the Frontiers for the Future project. We acknowledge the provision of computational facilities by the DJEI/DES/SFI/HEA Irish Centre for High-End Computing (ICHEC). MJK acknowledges the financial support from a Marie Sk\l odwoska-Curie Fellowship (Grant No. 101065974). JG is funded by a Science Foundation Ireland-Royal Society University Research Fellowship. This work was also supported by the European Research Council Starting Grant ODYSSEY (Grant Agreement No. 758403). MTM is supported by a Royal Society-Science Foundation Ireland University Research Fellowship (URF\textbackslash R1\textbackslash 221571). This project is co-funded by the European Union (Quantum Flagship project ASPECTS, Grant Agreement No. 101080167). Views and opinions expressed are however those of the authors only and do not necessarily reflect those of the European Union, Research Executive Agency or UKRI. Neither the European Union nor UKRI can be held responsible for them.

\end{acknowledgements}

\bibliography{stochastic_thermo_mesoleads.bib}
\bibliographystyle{apsrev4-1}
\clearpage
\appendix

\section{Connection between $C$ and $n_s$}
\label{sec: connection_C_n_s}
Given a fermionic Gaussian state $\rho$, of dimension $(2^N \times 2^N)$, with eigenstates $\lbrace \ket{s^o_k}\rbrace$ and eigenvalues $\lbrace\lambda_{s_k}\rbrace$, its covariance matrix $C$, of dimension $(N\times N)$, is
\begin{equation} \begin{split}
    C_{ij} &= \mathrm{Tr}\left[\rho c_j^ \dagger  c_i\right]\\
    &= \mathrm{Tr}\left[\sum_{k=1}^{2^N} \lambda_{s_k}\ket{s^o_k}\bra{s^o_k} c_j^ \dagger  c_i\right]\\
    &= \sum_{k=1}^{2^N}   \lambda_{s_k} \mathrm{Tr}\left[ \ket{s^o_k}\bra{s^o_k} c_j^ \dagger  c_i\right]\\
    &= \sum_{k=1}^{2^N}   \lambda_{s_k} C^{s_k}_{ij}.
 \end{split}\end{equation}
Let $U$ diagonalise $C$ such that
\begin{equation}
    C = U D U^ \dagger ,
\end{equation}
where $D$ is diagonal, with the eigenvalues of $C$ on its diagonal. 
We now denote
\begin{equation}
    C^{s_k} = U n_{s_k} U^ \dagger,
\end{equation}
where the set of matrices $\lbrace n_{s_k}\rbrace$ is simply given by the set of all diagonal matrices of size $(2^{N}\times 2^{N})$ with bit-strings of length $N$ (with letters either 0 or 1) on their diagonal.
We then expand the matrix elements of $C$ in terms of the matrices $C^{s_k}$
\begin{equation} \begin{split}
    C_{ij} &= \sum_{k=1}^{2^N} \lambda_{s_k} C^{s_k}_{ij}\\
    &= \sum_{k=1}^{2^N}  \lambda_{s_k} (U n_{s_k} U^ \dagger )_{ij}\\
    C &= U \left(\sum_{k=1}^{2^N} \lambda_{s_k} n_{s_k}\right) U^ \dagger .
 \end{split}\end{equation}
We can now identify $D = \sum_{k=1}^{2^N}  \lambda_{s_k} n_{s_k}$.

\section{Gaussianity of quantum-jump trajectories}
\label{sec:Gaussianity_appendix}

In this appendix, we show that the conditional quantum state remains Gaussian along the entire trajectory for the dynamics described in the main text, assuming the initial state is Gaussian. We invoke the results of Bravyi~\cite{bravyi_lagrangian_2005}, who provided a complete characterisation of Gaussian operators (e.g. Gaussian unitaries and Gaussian states) and Gaussian-preserving maps. In particular, we will use the fact that a product of Gaussian operators is itself Gaussian, and that projective measurements in the Fock basis preserve Gaussianity.

First we consider the no-jump evolution of an unnormalised state $\tilde{\rho}_r$, which is related to the normalised conditional state by $\rho_r = \tilde{\rho}_r/\mathrm{Tr}[\tilde{\rho}_r]$. If no jump is recorded, the unnormalised state evolves according to 
\begin{equation}
\label{unnormalised_evolution}
    \tilde{\rho}_r(t+{\rm d}t) = e^{-i H_{\rm eff} {\rm d}t} \tilde{\rho}_r(t) \left(e^{-i H_{\rm eff} {\rm d}t}\right)^\dagger,
\end{equation}
where the non-Hermitian Hamiltonian is
\begin{equation}
\label{H_eff}
    H_{\rm eff} = H - \frac{i}{2}\sum_{k=1}^L \sum_{ \sigma\in \lbrace +, - \rbrace}L_{k}^{\sigma\dagger} L_{k}^{\sigma}.
\end{equation}
After normalisation, Eq.~\eqref{unnormalised_evolution} is equivalent to the no-jump evolution in Eq.~\eqref{eq: rho_stochastic}, i.e. with ${\rm d}N_k^\sigma = 0$, where we retain terms up to first order in the infinitesimal time step ${\rm d}t$. At the same order, $e^{-i H_{\rm eff}{\rm d}t} = \Xi_{{\rm d}t} U_{{\rm d}t}$, where $U_{{\rm d}t}= e^{-i H {{\rm d}t}}$ is a unitary time evolution operator generated by a quadratic Hamiltonian, and  $ \Xi_{{\rm d}t} = e^{-\sum_{k,\sigma}L_k^{\sigma\dagger} L_k^
{\sigma}{\rm d}t} $ can be thought of as an (unnormalised) Gaussian density operator. Therefore, if $\tilde{\rho}_r(t)$ is Gaussian, Eq.~\eqref{unnormalised_evolution} is a product of Gaussian states and Gaussian unitaries. This proves that $\tilde{\rho}_t(t+{\rm d}t)$ is Gaussian and thus so is its normalised equivalent $\rho_r(t+{\rm d}t)$.

To prove that the jump evolution is Gaussian, we first note that the maps representing the state update after a projective measurement in the Fock basis,
\begin{equation}
\label{projection_maps}
    \mathcal{J}^-_k \rho  = \frac{c_k^\dagger c_k \rho c_k^\dagger c_k}{\mathrm{Tr}[c_k^\dagger c_k \rho ]}, \quad \mathcal{J}^+_k\rho = \frac{c_k c^\dagger_k \rho c_k c^\dagger_k}{\mathrm{Tr}[c_k c^\dagger_k \rho ]},
\end{equation}
are Gaussian-preserving maps~\cite{bravyi_lagrangian_2005}. Indeed, fermionic Fock states are connected to the vacuum by a Gaussian unitary, as discussed in Sec.~\ref{sec:computing_probs}. The maps~\eqref{projection_maps} are unitarily equivalent to the action of quantum jumps that create or destroy a fermion in mode $k$, since
\begin{align}
    & \frac{c_k \rho c_k^\dagger}{\mathrm{Tr}[c_k^\dagger c_k \rho ]} = U_{\rm ph} \left(\mathcal{J}^-_k\rho \right) U^\dagger_{\rm ph}, \notag \\ 
    & \frac{c^\dagger_k \rho c_k}{\mathrm{Tr}[c_k c^\dagger_k \rho ]} = U_{\rm ph} \left(\mathcal{J}^+_k\rho \right)  U^\dagger_{\rm ph},
\end{align}
where $U_{\rm ph} = \prod_{k}(c_k + c_k^\dagger)$ is a unitary particle-hole transformation. It is straightforward to prove that $U_{\rm ph}$ is a Gaussian unitary because it preserves the canonical anti-commutation relations. Its action simply swaps creation and annihilation operators, $U_{\rm ph}c_k U_{\rm ph}^\dagger = \pm c_k^\dagger$, up to an irrelevant phase factor. We thus see that the quantum jump evolution 
\begin{align}
    \rho_r(t+{\rm d}t) & = \rho_r(t) + \mathcal{G}[L_k^\pm] \rho_r(t) \notag \\ &= U_{\rm ph}\left(\mathcal{J}^\pm_k \rho_r(t)\right)U_{\rm ph}^\dagger
\end{align}
is the composition of a Gaussian-preserving projective measurement followed by a Gaussian unitary transformation, and therefore is itself a Gaussian-preserving map.

\section{Imperfect detection}
\label{sec: imperfect_detection}
In the following, we generalise the results presented in Secs.~\ref{sec: cond_evo} and ~\ref{sec: cond_thermo} to scenarios of imperfect measurement or different measurement configurations.
\subsection{Conditional dynamics}
\begin{figure*}[t]
\begin{center}
\includegraphics[width=\linewidth]{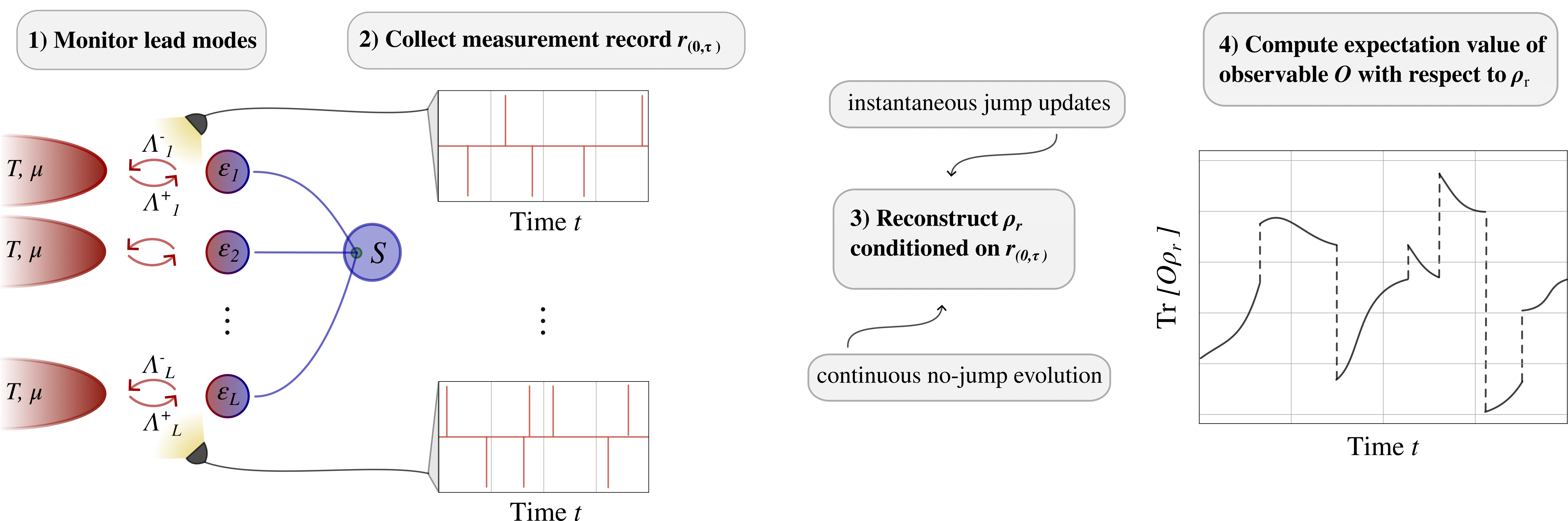}
  \caption{Particle transfer between all lead modes and their respective residual reservoirs is recorded with measurement efficiencies $\Lambda_k^\sigma$. The expectation value of an observable $O$ is computed with respect to the state conditioned on the measurement record $r$. When a quantum jump is recorded, the state undergoes an abrupt update, leading to an instantaneous change in the expectation value of the observable. During the intervals between jumps, the state evolves smoothly following the no-jump conditional evolution.}
 \label{fig: observable_trajectory_imperfect}
\end{center}
\end{figure*}
To be more general, we assume the freedom to independently choose to monitor the distinct jump channels with different efficiencies denoted as $\Lambda^\sigma_k$. Here, a perfect detection event is represented by $\Lambda^\sigma_k = 1$, while setting $\Lambda^\sigma_k = 0$ implies that the corresponding jump channel remains entirely unmonitored \cite{wiseman_quantum_2009}.
The stochastic ME then is given by
\begin{equation}
\begin{split}
    \mathrm{d}\rho_r =& -i\left[H, \rho_r\right]\mathrm{d}t-\frac{1}{2}\sum_{k=1}^L \sum_{ \sigma\in \lbrace +, - \rbrace} \mathcal{H}\left[{L^\sigma_k}^ \dagger  L^\sigma_k\right]\rho_r \mathrm{d}t\\
    &+ \sum_{k=1}^L \sum_{ \sigma\in \lbrace +, - \rbrace} (1- \Lambda_k^\sigma) \mathcal{G}\left[L^\sigma_k\right]\rho_r \mathrm{Tr}\left[{L^\sigma_k}^\dagger L^\sigma_k \rho_r\right]\mathrm{d}t\\
    &+ \sum_{k=1}^L \sum_{ \sigma\in \lbrace +, - \rbrace}   \mathcal{G}\left[L^\sigma_k\right]\rho_r \mathrm{d}N^\sigma_k,
\end{split}
\end{equation}
where $\mathrm{E}[\mathrm{d}N^\sigma_k] = \Lambda_k^\sigma \mathrm{Tr}\left[{L^\sigma_k}^\dagger L^\sigma_k \rho_r\right]\mathrm{d}t$.
It follows that the covariance matrix evolves, conditioned on the measurement record $r$, as
\begin{equation}
\label{eq: C_nojump}
\begin{split}
    \mathrm{d}C^r_{ij}(t)& 
    =  i\left\langle\left[H, c_j^\dagger c_i \right]\right\rangle_r \mathrm{d}t\\
  &-\frac{1}{2}\sum_{k=1}^L \sum_{ \sigma\in \lbrace +, - \rbrace}  \left\langle\mathcal{H}^+\left[{L^\sigma_k}^\dagger L^\sigma_k\right]c^ \dagger _j c_i \right\rangle_r\mathrm{d}t\\
     &+ \sum_{k=1}^L \sum_{ \sigma\in \lbrace +, - \rbrace} (1- \Lambda_k^\sigma) \left\langle\mathcal{G}^+\left[L^\sigma_k\right]c_j^\dagger c_i \right\rangle_r \langle {L^\sigma_k}^\dagger L^\sigma_k \rangle_r\mathrm{d}t\\
      &+ \sum_{k=1}^L \sum_{ \sigma\in \lbrace +, - \rbrace}   \left\langle\mathcal{G}^+\left[L^\sigma_k\right]c_j^\dagger c_i \right\rangle_r \mathrm{d}N_k^\sigma.
\end{split}
\end{equation}
\subsubsection{No jump-conditioned evolution}
Between two recorded jumps, $C^r$ evolves according to the matrix equation
\begin{equation}
\begin{split}
    \frac{\mathrm{d}C^r}{\mathrm{d}t} =& -(VC^r + C^r V^ \dagger ) + C^r B C^r \\
    &+ \bar{C}^r \left(F\left(1-\Lambda^+\right)\right)\bar{C}^r \\
    &- C^r \left((\Gamma - F)\left(1-\Lambda^-\right)\right)C^r,
\end{split}
\end{equation}
where $B = \Gamma - 2F$, $V = i H + \frac{1}{2}B = W-F$ and $\bar{C}^r= 1-C^r$. Further, $\left(1-\Lambda^{\sigma}\right)_{kk} = 1-\Lambda^{\sigma}_{kk}$.
In between jumps, the survival probability $p$ evolves under the differential equation
\begin{equation} \begin{split}
    \frac{\mathrm{d} p}{\mathrm{d}t} &= - p \sum_{k=1}^L \sum_{ \sigma\in \lbrace +, - \rbrace}  \Lambda_k^\sigma \left\langle {L^\sigma_k}^\dagger {L^\sigma_k} \right\rangle_r, \\
    \end{split}\end{equation} 
 where we have used that $p$ is given by the trace of the normalized density matrix between jumps.
Therefore, the survival probability decays as
\begin{equation}
\begin{split}
    p(t) &= p(t_{J_{m-1}})  \exp\left(-\int_{t_{J_{m-1}}}^{t} K(s)\mathrm{d} s\right),
\end{split}
\end{equation}
where
\begin{equation}
\begin{split}
    K(s) &=   \mathrm{Tr}\left[\Lambda^+F\bar{C}^r(s)\right] + \mathrm{Tr}\left[\Lambda^-(\Gamma-F)C^r(s)\right] .
\end{split}
\end{equation}\\
\subsubsection{Jump-conditioned evolution}
Upon recording a jump in a lead mode, denoted by $(t_{J_m}, k_{m}, \sigma_{m})$, $C^r$ is updated instantaneously. At $t = t_{J_m}$, $\mathrm{d}N_{k, \sigma}(t_{J_m})=\delta_{k k_{m}}\delta_{\sigma \sigma_{m}}\in \lbrace 0,1 \rbrace$ and therefore
\begin{equation}
\begin{split}
    \mathrm{d}C^r_{ij}(t_{J_m}) 
     =& \left\langle \mathcal{G}^+\left[{L^{\sigma_{m}}_{k_{m}}}\right]c^ \dagger _j c_i\right\rangle_r(t_{J_m}) .
\end{split}
\end{equation}
If a jump onto the lead mode $k_{m}$ from its residual reservoir is recorded, so that  $\sigma_{m} = +$, then
\begin{equation}
    \label{eq: jump_update_C_+_imperfect}
    \mathrm{d}C^r(t_{J_m}) =  \bar{C}^r \left(\frac{\mathrm{d}N_{+}(t_{J_m})}{\bar{C}^r}\right) \bar{C}^r,
\end{equation}
where $\bar{C}^r = 1-C^r$.
Otherwise, if one records a jump off the lead mode $k_{m}$ to its residual reservoir, so that  $\sigma_{m} = -$, then
\begin{equation}
      \mathrm{d}C^r(t_{J_m}) = - C^r \left(\frac{\mathrm{d}N^{-}(t_{J_m})}{C^r}\right) C^r.
\end{equation}
Above, we use the shorthand
\begin{equation}
    \begin{split}
        \left(\frac{\mathrm{d}N^{+}(t)}{\bar{C}^r} \right)_{kk} &= \frac{\mathrm{d}N^+_{k}(t)}{1-C^r_{kk}} 
        ,\\
\left(\frac{\mathrm{d}N^{-}(t)}{C^r}\right)_{kk} &=
    \frac{\mathrm{d}N^-_{k}(t)}{C^r_{kk}}
    .
    \end{split}
\end{equation}
\subsubsection{Stochastic ME for the covariance matrix}
The stochastic ME governing the trajectory of $C^r$, conditioned on the imperfect measurement record $r$, is given by
\begin{equation}
    \begin{split}
        \mathrm{d}C^r(t) = & \left[-(VC^r + C^r V^ \dagger ) + C^r B C^r\right]\mathrm{d}t\\
        &+ \bar{C}^r F\left(1-\Lambda^+\right)\bar{C}^r\mathrm{d}t \\
        &- C^r (\Gamma - F)\left(1-\Lambda^-\right)C^r\mathrm{d}t\\
        &  +\bar{C}^r \left(\frac{\mathrm{d}N^{+}(t)}{\bar{C}^r}\right) \bar{C}^r \\
        & - C^r \left(\frac{\mathrm{d}N^{-}(t)}{C^r}\right) C^r.
    \end{split}
\end{equation}
\subsection{Particle current}
The stochastic particle current conditioned on the imperfect measurement record $r$ into the extended system via lead $\alpha$ is given by
\begin{equation}
\begin{split}
     I^{r}_{N_\alpha}(t) \mathrm{dt} = & \mathrm{Tr}\left[C^r B_{\alpha}C^r - B_{\alpha}C^r\right] \mathrm{d}t \\
      &+  \mathrm{Tr}\left[ \bar{C}^r\left(1-\Lambda^+_\alpha\right) F_\alpha \bar{C}^r \right]\mathrm{d}t \\
    &-  \mathrm{Tr}\left[C^r\left(1-\Lambda^-_\alpha\right)(\Gamma_\alpha-F_\alpha)C^r\right]\mathrm{d}t\\
    &+  \mathrm{Tr}\left[ \bar{C}^r\left(\frac{\mathrm{d}N^+_{ {\alpha}}(t)}{\bar{C}^r}\right) \bar{C}^r \right] \\
    &-  \mathrm{Tr}\left[C^r\left(\frac{\mathrm{d}N^-_{{\alpha}}(t)}{C^r}\right)C^r\right],
\end{split}
\end{equation}
where $B_\alpha=\Gamma_\alpha - 2F_\alpha$ with
\begin{equation}
\begin{split}
    \left(\Gamma_\alpha \right)_{kk}, \left(F_\alpha \right)_{kk}= \begin{cases} 
      \Gamma_{kk}, F_{kk} &  \text{if $k$-mode in lead $\alpha$,} \\
      0 & \text{otherwise.}\\
   \end{cases}\\
    \left(1-\Lambda^\sigma_\alpha\right)_{kk}= \begin{cases} 
      1-\Lambda^\sigma_{kk} &  \text{if $k$-mode in lead $\alpha$,} \\
      0 & \text{otherwise.}\\
   \end{cases}
\end{split}
\end{equation}
We use the shorthand
\begin{equation}
    \begin{split}
        \left(\frac{\mathrm{d}N^+_{\alpha}(t)}{\bar{C}^r} \right)_{kk} &= \begin{cases}
    \frac{\mathrm{d}N^+_{k, \alpha}(t)}{\bar{C}^r_{kk}} & \text{ if $k$-mode in lead $\alpha$,} \\
    0 & \text{ otherwise.}
\end{cases}\\
\left(\frac{\mathrm{d}N^-_{\alpha}(t)}{C}\right)_{kk} &= \begin{cases}
    \frac{\mathrm{d}N^-_{k, \alpha}(t)}{C^r_{kk}} & \text{ if $k$-mode in lead $\alpha$,} \\
    0 & \text{ otherwise.}
\end{cases}
    \end{split}
\end{equation}
\subsection{Energy current}
The stochastic energy current conditioned on the imperfect measurement record $r$ into the extended system via lead $\alpha$ is given by
\begin{equation}
\begin{split}
     I^{r}_{E_\alpha}(t) \mathrm{d}t = & \ \mathrm{Tr}\left[B_\alpha\left(C^r HC^r - \frac{1}{2}(HC^r+C^r H)\right)\right] \mathrm{d}t \\
     &+  \mathrm{Tr}\left[H \bar{C}^r \left(1-\Lambda^+_\alpha\right)F_\alpha \bar{C}^r\right]\mathrm{d}t \\
    &-\mathrm{Tr}\left[HC^r\left(1-\Lambda^-_\alpha\right)(\Gamma_\alpha - F_\alpha)C^r\right]\mathrm{d}t\\
    &+  \mathrm{Tr}\left[ H \bar{C}^r\left(\frac{ \mathrm{d}N^+_\alpha(t)}{\bar{C}}\right)\bar{C}^r\right] \\
    &-\mathrm{Tr}\left[HC^r\left(\frac{\mathrm{d}N^-_\alpha(t)}{C^r}\right)C^r\right].
\end{split}
\end{equation}
\subsection{Measurement energy current}
The stochastic measurement energy current conditioned on the imperfect measurement record $r$ is defined as
\begin{widetext}
\begin{equation}
\begin{split}
    I_{E_\mathrm{M}}\mathrm{d}t = &-\frac{\mathrm{d}t}{2} \sum_{k=1}^L \sum_{\sigma \in \lbrace +, - \rbrace} \mathrm{Tr}\left[H\lbrace {L^\sigma_k}^\dagger L^\sigma_k, \rho_r\rbrace\right] 
    +\mathrm{d}t\sum_{k=1}^L \sum_{\sigma \in \lbrace +, - \rbrace}  \langle H \rangle_r \langle  {L^\sigma_k}^ \dagger  L^\sigma_k\rangle_r
    \\
    &
    +\sum_{k=1}^L \sum_{\sigma \in \lbrace +, - \rbrace}  \mathrm{d}N^\sigma_k \frac{\mathrm{Tr}\left[{L^\sigma_k}^\dagger H_\mathrm{int} L^\sigma_k \rho_r\right]}{\langle  {L^\sigma_k}^ \dagger  L^\sigma_k\rangle_r} 
     +\mathrm{d}t \sum_{k=1}^L \sum_{\sigma \in \lbrace +, - \rbrace} (1-\Lambda_k^\sigma) \mathrm{Tr}\left[{L^\sigma_k}^\dagger H_\mathrm{int} L^\sigma_k \rho_r\right] 
     \\
    &
    +\frac{1}{2}\sum_{k=1}^L \sum_{\sigma \in \lbrace +, - \rbrace} \mathrm{d}N^\sigma_k\frac{\mathrm{Tr}\left[\lbrace  {L^\sigma_k}^ \dagger  L^\sigma_k, H_0 \rbrace \rho_r\right]}{\langle  {L^\sigma_k}^ \dagger  L^\sigma_k\rangle_r} 
    -\sum_{k=1}^L \sum_{\sigma \in \lbrace +, - \rbrace} \mathrm{d}N^\sigma_k \langle  H\rangle_r\\
    &+\frac{\mathrm{d}t}{2} \sum_{k=1}^L \sum_{\sigma \in \lbrace +, - \rbrace} (1-\Lambda_k^\sigma)\mathrm{Tr}\left[H_0\lbrace {L^\sigma_k}^\dagger L^\sigma_k, \rho_r\rbrace\right]
    -\mathrm{d}t \sum_{k=1}^L \sum_{\sigma \in \lbrace +, - \rbrace} (1-\Lambda_k^\sigma) \langle H \rangle_r \langle  {L^\sigma_k}^ \dagger  L^\sigma_k\rangle_r .
\end{split}
\end{equation}
\end{widetext}
The measurement energy current for lead $\alpha$ can be expressed in terms of the covariance matrix $C^r$
\begin{widetext}
    \begin{equation}
\begin{split}
    I_{E_{\mathrm{M}, \alpha}}\mathrm{d}t
    = & \mathrm{Tr}\left[B_\alpha\left(C^r HC^r - \frac{1}{2}(HC^r+C^r H)\right)\right] \mathrm{d}t \\
&+\mathrm{Tr}\left[\left(1-\Lambda^+_\alpha\right)F_\alpha\bar{C}^rH_\mathrm{int}\bar{C}^r\right]\mathrm{d}t
    -\mathrm{Tr}\left[\left(1-\Lambda^-_\alpha\right)(\Gamma_\alpha - F_\alpha)C^r H_\mathrm{int} C^r\right]\mathrm{d}t\\
    & -\frac{1}{2}\mathrm{Tr}\left[\left(1-\Lambda^+_\alpha\right)F_\alpha \left(\bar{C}^r H_0 C^r + C^r H_0 \bar{C}^r\right)\right]\mathrm{d}t
     -\frac{1}{2}\mathrm{Tr}\left[\left(1-\Lambda^-_\alpha\right)(\Gamma_\alpha - F_\alpha)\left(\bar{C}^r H_0 C^r + C^r H_0 \bar{C}^r\right)\right]\mathrm{d}t\\
    &+\mathrm{Tr}\left[\left(\frac{\mathrm{d}N^+_{\alpha}}{\bar{C}^r}\right)\bar{C}^rH_\mathrm{int} \bar{C}^r\right]
    -\mathrm{Tr}\left[\left(\frac{\mathrm{d}N^-_\alpha}{C^r}\right)C^r H_\mathrm{int} C^r\right]\\
    & -\frac{1}{2}\mathrm{Tr}\left[\left(\frac{\mathrm{d}N^+_\alpha}{\bar{C}^r} \right)\left(\bar{C}^r H_0 C^r + C^r H_0 \bar{C}^r\right)\right]
     -\frac{1}{2}\mathrm{Tr}\left[\left(\frac{\mathrm{d}N^-_\alpha}{C^r}\right)\left(\bar{C}^r H_0 C^r + C^r H_0 \bar{C}^r\right)\right].
\end{split}
\end{equation}
\end{widetext}
\section{Monte Carlo trajectory sampling}
\label{sec: monte_carlo_solver}
To numerically generate quantum trajectories, we employ the following Monte Carlo algorithm:
\begin{itemize}
    \item draw a random number $R_1$ between 0 and 1 (uniformly distributed).
    \item time evolve the survival probability $p$ and the covariance matrix $C^r$, conditioned on the measurement record $r$,  for a duration of length $t_J$ such that $p(t_J) = R_1 $. 
    \item The probability for a jump prescribed by the collapse operator $L^\sigma_k$ after time $t_J$ to occur is given by
    \begin{equation}
        P_{k, \sigma}(t_J) = \frac{\langle  {L^\sigma_k}^ \dagger   L^\sigma_k \rangle_r(t_J)}{1-R_1}.
    \end{equation}
    \item draw a second random number $R_2$ between 0 and 1 (uniformly distributed) to determine which jump to perform:  The corresponding collapse operator $L^{\sigma_{m}}_{k_{m}}$ is the first one for which the ordered sum
 \begin{equation}
    \sum_{k=1}^{L} \sum_{ \sigma\in \lbrace +, - \rbrace}  P_{k,\sigma}(t_J)\geq R_2.
 \end{equation}
Note that by definition, $\sum_{k=1}^{L} \sum_{ \sigma\in \lbrace +, - \rbrace} P_{k, \sigma}(t_J) = 1$. Finally, the covariance matrix is updated according to the corresponding jump as detailed in Sec.~\ref{sec: cond_evo}.
\end{itemize}
Note that $P_{k, \sigma}(t_J)$ can be easily expressed in terms of the diagonal elements of the covariance matrix, since  $\langle  L_k^{+ \dagger } L^+_k \rangle_r=\gamma_k f_k (1-C^r_{kk})$ and $\langle  L_k^{- \dagger }  L^-_k \rangle_r=\gamma_k (1-f_k) C^r_{kk}$.
\section{Steady-state heat current and entropy production in local GKSL master equations}
\label{sec: FT_appendix}

\begin{figure*}[t]
\begin{center}
\includegraphics[width=\linewidth]{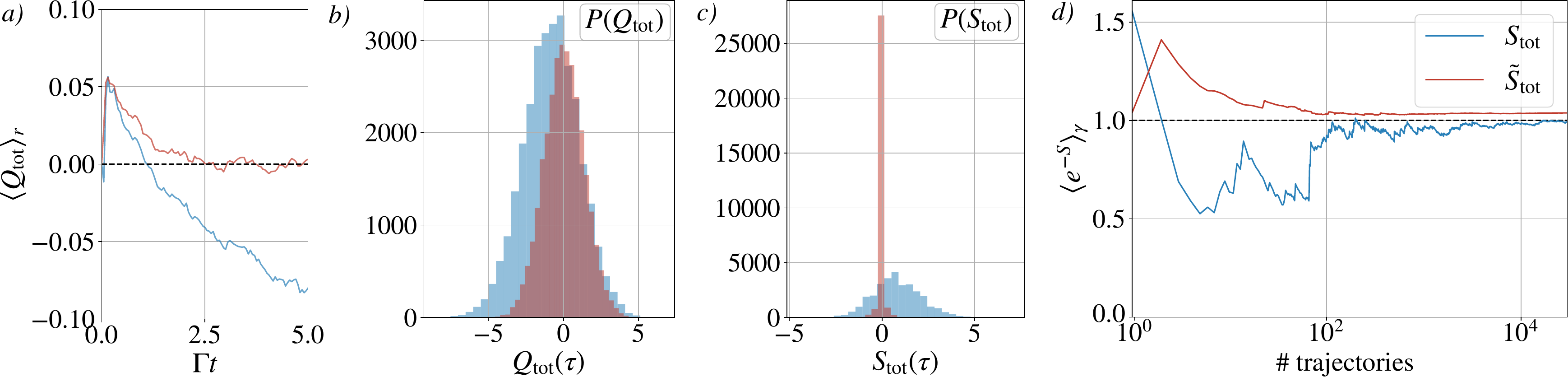}
  \caption{Time-evolving steady-state heat current and entropy production after the first projective measurement, shown for cases excluding (blue) and including (red) the measurement energy current. a) Time-evolving average of the integrated heat current. When the measurement energy current is excluded, a spurious heat current persists even in the steady state. Including the measurement energy current corrects this discrepancy. b) Distribution of the integrated heat current. c) Distribution of the total entropy production. The modified entropy production, $\tilde{S}_\mathrm{tot}$, is centered and sharply peaked around zero, as expected in a steady state. d) Convergence to the integral fluctuation theorem (Eq.~\eqref{eq: integral_FT}) for $S_\mathrm{tot}$, but not for $\tilde{S}_\mathrm{tot}$. Parameters:
  $L$ = 10, up to 30000 trajectories, $\omega_\mathrm{max} = 1$, $\epsilon$ = $\omega_\mathrm{max}/4$, $\Gamma = \omega_\mathrm{max}/8$, $T =\omega_\mathrm{max}$, $\mu = \omega_\mathrm{max}/16$, $\Gamma\tau = 50$.
}
 \label{fig: steadystate_EP_local_GKSL}
\end{center}
\end{figure*}

In the following, we elaborate on why, in the mesoscopic-leads formalism or any local GKSL master equation, entropy production along trajectories remains nonzero in a steady state and why its mean is nonzero. This arises from the definition of total entropy production
\begin{equation}
    S_\mathrm{tot}(r_{\left[0, \tau\right]}) = \log\left(\frac{p^0_{n}}{p^\tau_m}\right) + \Sigma_{r_{\left(0, \tau\right)}},
\end{equation}
where $\Sigma_{r_{\left(0, \tau\right)}} = -\frac{\Delta Q_r}{T}$ and 
\begin{equation}
    \Delta Q_r = \sum_{\alpha=1}^{N_\mathrm{R}} \sum_{k=1}^{l_\alpha} \sum_{\sigma \in \lbrace+,-\rbrace}  \int_0^\tau \frac{\Delta E^\sigma_{k,\alpha} - \mu_\alpha \Delta N^\sigma_{k,\alpha}}{T_\alpha}  \mathrm{d}N_k(t).
\end{equation}
Assume the initial state is the steady state, and the first projective measurement yields outcome $n_0$, corresponding to the steady-state eigenstate projector $\Pi_{n_0}$. This state then evolves stochastically over time $\tau$. Crucially, the state at the end of this trajectory is not guaranteed to be a steady-state eigenstate. Instead, it has finite overlap with a set of steady-state eigenstates (typically of size $>1$) and may be projected into any of these by the second measurement in the two-point measurement (TPM) scheme. 
As a result, the relative surprisal between the two measurement outcomes cannot be uniquely associated with a specific entropy flux $\Sigma_{r_{\left(0, \tau\right)}}$. Consequently, even in a steady state, there exists a distribution of total entropy flux. The nonzero average arises because measurement energy is excluded from Eq.~\eqref{eq: S_tot}, as illustrated in Fig.~\ref{fig: steadystate_EP_local_GKSL} a)–c). Including the measurement energy yields a modified entropy production $\tilde{S}_{\mathrm{tot}}$ which, however, would violate the integral fluctuation theorem in Eq.~\eqref{eq: integral_FT}~\cite{manzano_quantum_2022}, as shown in Fig.~\ref{fig: steadystate_EP_local_GKSL} d).
Interestingly, while this modified entropy production appears to violate the standard fluctuation theorem, it seems to converge rapidly to some fixed, finite value. Exploring this feature further would be an intriguing direction for future work, but it lies beyond the scope of this study.
The discrepancy between $S_\mathrm{tot}$ and $\tilde{S}_\mathrm{tot}$ is expected to be suppressed with an increasing number of lead modes per bath, $l$, as the system-lead coupling, which contributes to the measurement energy current, scales as $\kappa_{k} \propto 1/\sqrt{l}$. However, due to both statistical constraints and the computational cost of determining $p_m^{r\tau}$ (see Sec.~\ref{sec: EP_TPM}), which grows exponentially with the size of the extended system even for Gaussian systems, we limited our analysis to $l= 10$, given the available computational resources. Since sampling the distribution of the (integrated) energy and particle currents, as opposed to the entropy production, does not require relying on the TPM, leads with a larger number of modes become feasible, as demonstrated in Sec.~\ref{sec: heat_ft_le}, where $l = 20$ .
\end{document}